\documentclass[12pt]{article}
\usepackage{jheppub} 
\pdfoutput=1 
\usepackage{amsfonts,amsmath,amssymb,epsf,physics}
\usepackage{amsmath,braket}
\usepackage{subcaption}
\graphicspath{{figures/} }
\usepackage{graphicx,hyperref,color}
\hypersetup{
    bookmarks=true,         
    unicode=false,          
    pdftoolbar=true,        
    pdfmenubar=true,        
    linktocpage=true,       
    pdffitwindow=false,     
    pdfstartview={FitH},    
    pdfnewwindow=true,      
    colorlinks=true,       
    linkcolor=blue,          
    citecolor=blue,        
    filecolor=blue,      
    urlcolor=blue           
}
\usepackage{tensor}     
\usepackage{romannum}
\usepackage[dvipsnames]{xcolor}

\usepackage{lscape}

\numberwithin{equation}{section}									

\usepackage{comment}

\newcommand{\be}{\begin{equation}}
\newcommand{\ba}{\begin{eqnarray}}
\newcommand{\ea}{\end{eqnarray}}
\newcommand{\ee}{\end{equation}}

\newcommand{\s}{\sqrt}

\newcommand{\bea}{\begin{eqnarray}}
\newcommand{\eea}{\end{eqnarray}}
\newcommand{\bes}{\begin{equation*}}
\newcommand{\beas}{\begin{eqnarray*}}
\newcommand{\eeas}{\end{eqnarray*}}
\newcommand{\bas}{\begin{array*}}
\newcommand{\eas}{\end{array*}}
\newcommand{\ees}{\end{equation*}}

\newcommand{\ov}{\overline}

\title{\boldmath Thermal Pseudo-Entropy}
\author[a,b,d]{Pawel Caputa,}
\author[b,c]{Bowen Chen,}
\author[d,e]{Tadashi Takayanagi,}
\author[d]{Takashi Tsuda}

\affiliation[a]{The Oscar Klein Centre and Department of Physics, Stockholm University, AlbaNova, 106 91 Stockholm, Sweden}
\affiliation[b]{Faculty of Physics, University of Warsaw, Pasteura 5, 02-093 Warsaw, Poland}
\affiliation[c]{Kavli Institute for Theoretical Sciences, University of Chinese Academy of Sciences, Beijing 100190, China}
\affiliation[d]{Center for Gravitational Physics and Quantum Information, Yukawa Institute for Theoretical Physics, Kyoto University, Kitashirakawa Oiwakecho, Sakyo-ku, Kyoto 606-8502, Japan}
\affiliation[e]{Inamori Research Institute for Science, 620 Suiginya-cho, Shimogyo-ku,Kyoto 600-8411 Japan}

\emailAdd{pawel.caputa@fysik.su.se}
\emailAdd{chenbowen@ucas.ac.cn}
\emailAdd{takayana@yukawa.kyoto-u.ac.jp}
\emailAdd{takashi.tsuda@yukawa.kyoto-u.ac.jp}

\abstract{In this work, we develop a generalisation of the thermal entropy to complex inverse temperatures, which we call the thermal pseudo-entropy. We show that this quantity represents the pseudo-entropy of the transition matrix between Thermofield Double states at different times. We have studied its properties in various quantum mechanical setups, Schwarzian theory, Random Matrix Theories, and 2D CFTs, including symmetric orbifolds. Our findings indicate a close relationship between the averaged thermal pseudo-entropy and the spectral form factor, which is instrumental in distinguishing chaotic and integrable models. Moreover, we have observed a logarithmic scaling of this quantity in models with a continuous spectrum, with a universal coefficient that is sensitive to the scaling of the density of states near the edge of the spectrum. Lastly, we found the connection between the real and imaginary parts of the thermal pseudo-entropy through the Kramers-Kronig relations.}

\begin{document}
\begin{flushright}
YITP-24-151
\end{flushright}
	\maketitle
	\flushbottom

\section{Introduction}
Recently an interesting generalisation of entanglement entropy \cite{Bombelli:1986rw,Srednicki:1993im,Holzhey:1994we,Calabrese:2004eu,Calabrese:2009qy} has been introduced by considering post-selection processes in quantum many-body systems, which is called pseudo-entropy \cite{Nakata:2021ubr}. This quantity is defined from the transition matrix, which depends on two different states (i.e. the initial state and final state), while the density matrix depends only on a single state. Since the transition matrix is non-Hermitian, the pseudo-entropy becomes complex-valued in general. 

This quantity enjoys a number of useful properties. First of all, it has a simple and clear gravity dual in AdS/CFT correspondence \cite{Maldacena:1997re}, such that it can be directly computed from the area of the minimal area in a Euclidean time-dependent AdS background \cite{Nakata:2021ubr}. This provides a simple extension of holographic entanglement entropy \cite{Ryu:2006bv,Ryu:2006ef,Hubeny:2007xt} in AdS/CFT. It is also useful when we consider a time-like extension of entanglement entropy, the so called time-like entanglement entropy, in Lorentzian AdS spacetimes \cite{Doi:2022iyj,Liu:2022ugc,Doi:2023zaf,Guo:2024lrr,Anegawa:2024kdj,Heller:2024whi,Afrasiar:2024lsi}. Moreover, pseudo-entropy turns out to be a very useful probe for understanding the dS/CFT duality \cite{Strominger:2001pn,Maldacena:2002vr,Hikida:2021ese} as it computes the length of time-like geodesics \cite{Hikida:2022ltr,Narayan:2022afv,Doi:2023zaf,Narayan:2023ebn,Kawamoto:2023nki,Goswami:2024vfl}. For further progress of pseudo-entropy in the context of holography, field theories, and quantum information, refer to \cite{Miyaji:2021lcq,Akal:2021dqt,Mukherjee:2022jac,Guo:2022sfl,Ishiyama:2022odv,Miyaji:2022dna,Bhattacharya:2022wlp,Guo:2022jzs,Parzygnat:2022pax,He:2023eap,Kanda:2023zse,Xian:2023zgu,Parzygnat:2023avh,Guo:2023aio,Balasubramanian:2023xyd,Kanda:2023jyi,Kawamoto:2023ade,Wei:2024zez,Doi:2024nty,Parzygnat:2024tdf,Soni:2024aop,Hao:2024nhd,Guo:2024pve,fareghbal2024flat,jena2024note}.

In addition, we can apply pseudo-entropy to study quantum phase transitions in various quantum many-body systems. In particular, an intriguing quantity in this context is the difference between the real part of pseudo-entropy for two different states and the averaged entanglement entropy for each state. This quantity becomes non-positive when two states are in the same quantum phase. However, this difference can be positive when they are in the different phases as confirmed explicitly in \cite{Mollabashi:2020yie,Mollabashi:2021xsd}. It is also expected to be related to multi-partite entanglement \cite{Shinmyo:2023eci}.
Finally, pseudo-entropy in topological phases provides a new class of interesting topological measures \cite{Nishioka:2021cxe,Caputa:2024qkk} which can play roles of new order parameters.
Refer also \cite{Murciano:2021dga,Carignano:2023xbz,Carignano:2023xbz,
Carignano:2024jxb,Bou-Comas:2024pxf} for other condensed matter aspects of pseudo-entropy and related quantities.

One of the problems on pseudo-entropy, which we do not understand well at present, is the meaning of its imaginary part. In previous applications to quantum phase transitions, we have some qualitative understanding of the real part as mentioned above. However, so far, we have not been able to relate the imaginary part to any physical quantities such as e.g. order parameters. A quantum information theoretic meaning in terms of entanglement distillation in the presence of post-selection is only available for the real part of pseudo-entropy in a special class of quantum states \cite{Nakata:2021ubr}. Interestingly, in the context of dS/CFT, the imaginary part of the pseudo-entropy is expected to shed light on the emergence of time coordinate from a purely Euclidean CFT 
\cite{Hikida:2022ltr,Doi:2022iyj,Doi:2023zaf}.

In this paper, motivated by the above, we analyse the properties of the full complex-valued pseudo-entropy. To make the problem tractable, we focus on a simpler version of pseudo-entropy, which we call thermal pseudo-entropy (TPE). This is analogous to the relation between entanglement entropy and the standard thermal entropy, given by the statement that the entanglement entropy of a single copy in the Thermofield Double (TFD) state is simply equal to the thermal entropy. In other words, if we write the thermal entropy $S(\beta)$ as a function of the inverse temperature $\beta$, we define the thermal pseudo-entropy by the complex extension $S(\beta+it)$. We will see below that this precisely coincides with the pseudo-entropy for the TFD state and its time-evolution. 

This definition, by extending the standard thermal entropy, might be analogous to the spectral form factor (SFF) \cite{Cotler:2016fpe}. Even though they are different in that the former is a holomorphic quantity,
while the latter not, we will show that there is a useful relationship between them when we take time averages. For earlier works on this connection, refer also to \cite{Goto:2021kln,He:2024jog} that focused more on sub-regions and two-point correlators. Moreover, as we will see later, the imaginary part of TPE is completely determined by the real part via the Kramers-Kronig relations (and vice versa). This provides the first example with a sharp understanding of the imaginary part. We will work out several interesting properties of TPE by computing it in various examples below. This will also reveal how its time evolution depends on the energy spectra, taking into account the finite size effects.

This paper is organised as follows. In section \ref{sec:tpe}, we provide the definition and basic properties of TPE. In section \ref{sec:GeneralProp}, we explore some of its general properties, and in section \ref{sec:Examples} we analyse explicit examples such as harmonic oscillator, Calogero-Sutherland and Schwarzian models as well as Random Matrix Theories. In section \ref{sec:2dCFTs}, we present calculation of TPE in various two-dimensional CFTs.
In section \ref{sec:Orbifold}, we study the SFF and TPE in orbifold CFTs. In section \ref{sec:Concldis} we summarise and conclude.
In appendix \ref{sec:CompactScalarSymOrbFig}, we show plots of SFF and TPE for compact scalar symmetric orbifold CFTs.
\section{Thermal pseudo-entropy}\label{sec:tpe}
We start by briefly reviewing the notion of pseudo-entropy \cite{Nakata:2021ubr}, define the thermal pseudo-entropy and connect its averaging to the SFF. 
\subsection{Pseudo-entropy}
Pseudo-entropy is a generalisation of von Neumann entropy that involves the transition matrix between two quantum states $\ket{\psi}$ and $\ket{\phi}$ with non-zero overlap
\be
\tau^{\phi|\psi}=\frac{\ket{\phi}\bra{\psi}}{\langle \psi|\phi\rangle}. \label{tran-phi-psi}
\ee
As for ordinary entanglement entropy, we assume that Hilbert space admits a decomposition to a region $A$ and its complement $B$: $\mathcal{H}=\mathcal{H}_A\otimes \mathcal{H}_B$, and define the reduced transition matrix in analogy with reduced density matrix
\be
\tau^{\phi|\psi}_A=\Tr_B(\tau^{\phi|\psi}).
\ee
The pseudo-entropy is then defined as the von Neumann entropy of the reduced transition matrix 
\be
S(\tau^{\phi|\psi}_{A})=-\Tr(\tau^{\phi|\psi}_A\log \tau^{\phi|\psi}_A). \label{def:PE}
\ee
In the following, we will also consider the R\'enyi version of pseudo-entropy for an integer $n$
\be
S^{(n)}(\tau^{\phi|\psi}_{A})=\frac{1}{1-n}\Tr\left[(\tau^{\phi|\psi}_A)^n\right]. \label{def:PRE}
\ee
Since both quantities are based on the non-Hermitian transition matrix, in general, they are complex generalisations of von-Neumann and R\'enyi entropies. More detailed and pedagogical introduction to pseudo-entropy can be found in \cite{Nakata:2021ubr,Mollabashi:2021xsd,Parzygnat:2023avh}.
\subsection{Thermal pseudo-entropy}
In the remaining part of this work, we focus on a particular choice for the two states  $\ket{\psi}$ and $\ket{\phi}$ above. Namely, the first state will be the TFD state \cite{Takahashi:1996zn}, that is the canonical purification of the thermal density matrix with energy eigenvalues $E_n$ and inverse temperature $\beta=T^{-1}$ 
\be
\ket{\Psi_\beta}=\frac{1}{\sqrt{Z(\beta)}}\sum_n e^{-\frac{\beta}{2} E_n}\ket{E_n}_L\otimes\ket{E_n}_R.
\ee
The state requires two copies of the Hilbert space that we refer to as left (L) and right (R). The overall normalisation is given by the square root of the thermal partition function
\be
Z(\beta)=\sum_n e^{-\beta E_n}.
\ee
The second state will be the time evolution of the TFD with one of the Hamiltonians, say the left one, defined as
\be
\ket{\Psi_\beta(t)}=e^{-iH_Lt}\ket{\Psi_\beta}=\frac{1}{\sqrt{Z(\beta)}}\sum_n e^{-\frac{\beta+2it}{2} E_n}\ket{E_n}_L\otimes\ket{E_n}_R.
\ee
Next, we can define the transition matrix\footnote{Since we will only use the same two states throughout the paper, we drop the super-scripts from transition matrices.}
\be
\tau=\frac{\ket{\Psi_\beta(t)}\bra{\Psi_\beta}}{\langle \Psi_\beta|\Psi_\beta(t)\rangle},
\ee
with the overlap expressed by analytically continued partition function to complex inverse temperatures
\be
\langle \Psi_\beta|\Psi_\beta(t)\rangle=\frac{Z(\beta+it)}{Z(\beta)}.
\ee
The reduced transition matrix of the left subsystem becomes
\be
\tau_L=\Tr_R\left(\tau\right)=\frac{1}{Z(\beta+it)}\sum_n e^{-(\beta+it)E_n}\ketbra{E_n}.
\ee
This is simply the thermal density matrix with inverse temperature continued to complex values $\beta\to\beta+it$. \\
Finally, we define the thermal pseudo-entropy (TPE) as von-Neumann entropy of $\tau_L$
\be
S(\tau_L)=-\Tr(\tau_L\log \tau_L). \label{def:TPE}
\ee
The name TPE is clear since it coincides with the canonical thermal entropy
\be
S_{th}(\beta)=(1-\beta\partial_\beta)\log Z(\beta),\label{thent}
\ee
analytically continued to complex $\beta\to\beta+it$, which at $t=0$ reduces to the thermal entropy.\\
One intuitive way to rewrite the TPE is in terms of the complex phases of the transition matrix as
\be
S(\tau_L)=\log(Z(\beta+it))+\langle \theta(t)\rangle,
\ee
where
\be
\theta_n(t)\equiv(\beta+it)E_n,\qquad \langle \theta(t)\rangle=\frac{1}{Z(\beta+it)}\sum_n \theta_n(t)e^{-\theta_n(t)}. 
\ee
This also gives an interesting perspective that could be related to general ``complexity" measures designed to probe late time information hidden in the phases. We will not explore this direction here but hope to return to it in the near future.\\
In what follows, we will also be interested in R\'enyi versions of the thermal pseudo-entropies defined as
\be
S^{(n)}(\beta+it)=\frac{1}{1-n}\log\left[\frac{Z(n\beta)}{Z(\beta)^n}\right]_{\beta\to \beta+it},
\ee
that reduce to TPE in the $n\to 1$ limit.
\subsection{Averaged thermal pseudo-entropy and the spectral form factor}
Interestingly, we can write the real and imaginary parts of the thermal pseudo R\'enyi entropies as
\be
2(1-n)S^{(n)}(\beta+it)=\log\left[\frac{|Z(n(\beta+it))|^2}{|Z(\beta+it)|^{2n}}\right]+\frac{1}{i}\log\left[\frac{Z(n(\beta+it))Z(\beta-it)^n}{Z(\beta+it)^nZ(n(\beta-it))}\right].
\ee
Note that, in general, both real and imaginary parts are highly oscillating functions of $t$.
This also shows that we can think about the TPE as a close cousin of the SFF (refer also to \cite{Goto:2021kln}) and analogous probes that require analytic continuation of the real inverse temperatures $\beta$ into complex domain $\beta+it$. 

In fact, if we average over the energy spectrum, taking into account only two body correlations, the real part of the TPE reads (see also \cite{He:2024jog})
\ba
\ov{\mbox{Re}\,S^{(n)}(\beta+it)}=\frac{1}{2(1-n)}\left[\log \ov{|Z\left(n(\beta+it)\right)|^2}-n\log \ov{|Z\left(\beta+it\right)|^2}\right].
\ea
It is then clear that the real part can be related to the SFF \cite{Cotler:2016fpe}, defined by 
\ba
g_\beta(t)=\left|\frac{Z(\beta+it)}{Z(\beta)}\right|^2,
\ea
as follows
\ba
\ov{\mbox{Re}\,S^{(n)}(\beta+it)}=\frac{1}{2(1-n)}\log\frac{g_{n\beta}(t)}{g_\beta(t)^n}+S^{(n)}_{th}(\beta).  \label{SFFTPE}
\ea
We can make use of this connection to observe the following universal feature. Namely, if there is no degeneracy in the energy spectrum, the late time limit leads to
\ba
\ov{\mbox{Re}\,S^{(n)}(\beta+it)}
 \xrightarrow[t\to\infty]{}  \frac{1}{2(1-n)}\log \frac{Z(2n\beta)}{Z(2\beta)^n}= \frac{1}{2}S^{(n)}(2\beta). \label{LWW}
\ea
If there are degeneracies, we need to replace $Z(2n\beta)$ and $Z(2\beta)$ in (\ref{LWW}) with $\sum_{n}d_ne^{-2n\beta E_n}$ and $\sum_{n}d_ne^{-2\beta E_n}$, respectively, where $d_n\geq 1$ is the degree of degeneracy. It is then a simple exercise to show that, in general $\ov{\mbox{Re}\,S^{(n)}(\beta+it)}\geq\frac{1}{2}S^{(n)}(2\beta)$ .
\section{General properties of thermal pseudo-entropy}\label{sec:GeneralProp}
In this section we describe two important properties of TPE. The first is the Kramers-Kronig relation between its real and imaginary parts. The second is the late time behaviour in cases where we can approximate (``coarse-grain") the energy spectrum by a continuous function. 
\subsection{Kramers-Kronig relations}\label{sec:KK}
One of the most interesting and urgent questions regarding the pseudo-entropy in general are the meaning of its imaginary part and a potential relation with the real part in various setups. It turns out that, for the TPE, there is more mathematical structure that allows for progress and better understanding of these manners. This can be seen as follows.

Assuming that the function $f(t)$ is analytic in the upper half plane and decays rapidly in the limit Im$(t)\to\ -\infty$, we have the following relation between the real and imaginary parts
\ba
\mbox{Im}[f(t)]=\frac{1}{\pi}P\int^\infty_{-\infty}ds \frac{{\mbox{Re}}[f(s)]}{s-t},
\ea
where $P$ is the Cauchy principal value of the integral. This is the Hilbert transform, often referred to in physics as the Kramers–Kronig relations \cite{kramers1927diffusion,de1926theory} (KK-relations for short).\\
We can confirm this relation for TPE as a function of $t$, i.e. by taking $f(t)=S(\beta+it)$. The decay condition in the limit  Im$(t)\to\ -\infty$ is satisfied because the entropy $S(\beta)$ should go to zero in the zero temperature limit $\beta\to\infty$. Also, it is clear that the pseudo-entropy $S(\beta+it)$ is analytic in the upper half plane Im$(t)>0$. Indeed, we will explicitly confirm, by numerical integration, that the Kramers-Kronig relation holds in all examples considered in this paper.
\subsection{Time evolution under continuous spectrum approximation}
Next, we assume that we can approximate the energy spectrum by a continuous function. This is sometimes understood as a coarse-graining procedure in some limit of parameters of the model (e.g. large rank $N$ of a matrix). This way, we can simply write the density of states as $\rho(E)$ and assume the lowest energy denoted by $E_0$. The partition function is then computed as
\ba
Z(\beta)=\int^\infty_{E_0} dE e^{-\beta E} \rho(E).
\ea
If we assume the behaviour of $\rho(E)$ near the edge of the spectrum $E=E_0$ as follows
\ba
\rho(E)\sim (E-E_0)^\gamma,  \label{gammandef}
\ea
then we find, in the low temperature limit $\beta\to\infty$, the scaling of the partition function
\ba
Z(\beta)\sim \beta^{-1-\gamma}e^{-\beta E_0}.
\ea
This leads to the following behaviour of the thermal entropy (\ref{thent})
\ba 
S_{th}(\beta)\simeq -(1+\gamma)\log \beta,  \label{logS}
\ea
for large $\beta$. We will make use of this relation momentarily. However, we should keep in mind that, in the strict $\beta\to\infty$ limit of standard quantum systems, the discrete nature of energy spectrum becomes important and the behaviour (\ref{logS}) must be modified such that $\lim_{\beta\to\infty}S_{th}(\beta)=0$.

Now we turn to the TPE, which is obtained from the thermal entropy by replacing $\beta$ with $\beta+it$. As a consequence of \eqref{logS}, for large $t$, we have
\ba
\ov{S(\beta+it)}\simeq -(1+\gamma)\log t.  \label{logpe}
\ea
Note that $S(\beta+it)$ is in general a highly oscillating function and we need to average it 
over a small time window to find the behaviour (\ref{logpe}). Again, this logarithmic scaling occurs when we can approximate the energy spectrum by a continuous distribution. Thus the behaviour (\ref{logS}) and (\ref{logpe}) is valid only when $t\ll \frac{1}{\Delta E}$, where $\Delta E$ is the energy level spacing. We will show this in explicit examples below.

Moreover, from our general discussion, we expect that, as a function of $t$,  $\ov{S(\beta+it)}$ starts from the initial value $S_{th}(\beta)$ at $t=0$ and eventually gets decreased (plateau) to a positive value $S_*$, which satisfies 
\ba
S_*\geq \frac{1}{2}S^{(n)}(2\beta),  \label{SSW}
\ea
as we have seen in (\ref{LWW}). This allows us to estimate the maximal time $t_*$ where we can trust the behaviour (\ref{logpe}) as 
\be
(1+\gamma)\log(t_*)\sim S_{th}(\beta)-S_{*}.\ \ \ 
\ee
Next, to gain more intuition for different evolution scenarios of TPE, we turn to several explicit examples.
\section{Quantum mechanical examples}\label{sec:Examples}
In this section we analyse a few explicit examples where TPE can be computed. We also verify the KK relations between its real and imaginary parts and discuss the logarithmic contribution for the continuous spectrum models.
\subsection{Two-level systems}
Arguably one of the simplest models to start with is the two-level system or a particle with energy levels  $E_1=0$ and $E_2=\epsilon$. We can consider N non-interacting particles of this kind with partition function
\be
Z(\beta)=\left(\sum^2_{n=1}e^{-\beta E_n}\right)^N=2^N e^{-N\beta \epsilon/2}\cosh^N(\beta\epsilon/2).
\ee
For the reference, we can easily evaluate the SFF in this model that is periodic with period $t=2\pi/\epsilon$, as it should for this simple integrable model
\be
\left|\frac{Z(\beta+it)}{Z(\beta)}\right|^2=\left(\frac{\cosh(\beta\epsilon)+\cos(t\epsilon)}{\cosh(\beta\epsilon)+1}\right)^N.
\ee
The standard thermal entropy computed from \eqref{thent} becomes
\be
S_{th}(\beta)=N\log\left(1+e^{-\beta \epsilon}\right)+\frac{N\beta\epsilon}{1+e^{\beta\epsilon}}.
\ee
\begin{figure}[b!]
    \centering
    \includegraphics[width=10cm]{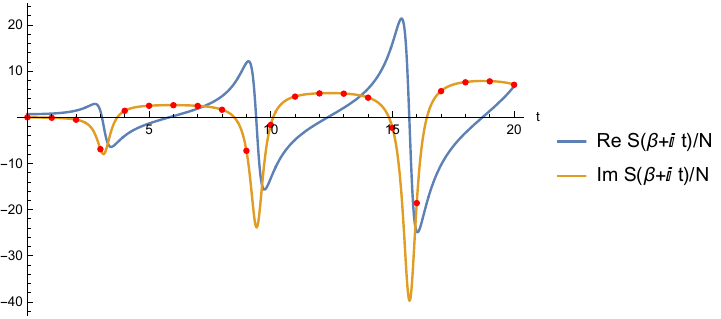}
    \caption{Real and imaginary parts of TPE \eqref{TPE2Level} for $\epsilon=1$ and $\beta=1/3$. Red dots indicate numerically computed imaginary part using the KK-relations.}
    \label{2Level}
\end{figure}
Finally, the TPE obtained by sending $\beta\to\beta+it$ becomes 
\be
S(\beta+it)=N\log\left(1+e^{-(\beta+it) \epsilon}\right)+\frac{N(\beta+it)\epsilon}{1+e^{(\beta+it)\epsilon}}.\label{TPE2Level}
\ee
We plotted its real and imaginary parts on Fig.\,\ref{2Level}. Both parts show periodic evolution with amplitudes that keep increasing in time. Moreover, we can verify that imaginary parts are related to the real part by the Kramers-Kroning relation. Numerical integral using the KK formula with the real part is marked by the red dots on the orange imaginary part for several instances of time.  
\subsection{Harmonic oscillator and Calogero-Sutherland model}
Next we consider the quantum harmonic oscillator and the TFD state of two copies of infinite-dimensional Hilbert space with energy spectrum and thermal partition function given by
\be
E_n=\omega (n+1/2),\qquad Z(\beta)=\frac{1}{2\sinh\left(\frac{\beta \omega}{2}\right)}.
\ee
Following the same steps as before, we derive the TPE which is now given by
\be
S_1(\beta+it,\omega)\equiv-\log\left[2\sinh\left(\frac{(\beta+it) \omega}{2}\right)\right]+\frac{(\beta+it)\omega}{2}\coth\left[\frac{(\beta+it)\omega}{2}\right].\label{TPEHO}
\ee
\begin{figure}[b!]
    \centering
    \includegraphics[width=7.5cm]{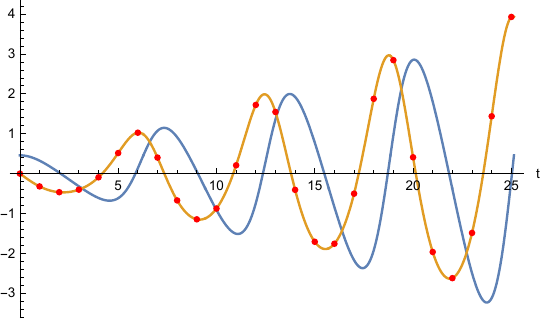}
    \hfill
    \includegraphics[width=7.5cm]{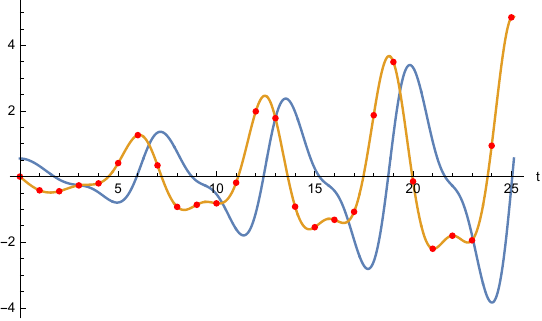}
    \caption{(Left) Real (blue) and imaginary (orange) parts of TPE for $\beta = 2$ and $\omega=1$ for the harmonic oscillator $N = 1$. (Right) For the Calorego-Sutherland model with $N=2$ and the same parameters. Red dots are numerical checks of the KK relations at several instances of time. }
    \label{fig:CSM}
\end{figure}
We present the real and imaginary parts on Fig.\,\ref{fig:CSM} (left). They oscillate in time with period determined by $\omega$. The amplitude of these oscillations keeps increasing with time. This can be seen from the linear $t$ term in the part of expression \eqref{TPEHO}. Somewhat surprisingly, this linear growth of the amplitude of TPE with time is a feature of integrable models as well as of chaotic ones, as we will see later.\\
Note that the partition function is exponentially suppressed for $\beta\to\infty$. Hence, the KK relations are also satisfied and we present their verification by red dots on the imaginary part on Fig.\,\ref{fig:CSM}.

A closely related generalisation is the famous integrable Hamiltonian of the Calorego-Sutherland model (CSM)
\be
H=\frac{1}{2}\sum^N_{i=1}\left[-\frac{\partial^2}{\partial z^2_i}+\omega^2 z^2_i\right]+\sum_{i<j}\frac{\lambda(\lambda-1)}{(z_i-z_j)^2}.
\ee
The SFF for this example was already discussed in \cite{delCampo:2017bzr}. It turns out, that the coupling $\lambda$ only renormalises the vacuum energy and the partition function of the model can be written as a product over partition functions of a single harmonic oscillator with different frequencies as follows
\be
Z(\beta)=\prod^N_{k=1}\frac{1}{2\sinh\left(k\frac{\beta\omega}{2}\right)}.
\ee
The SFF is periodic in time and given by \cite{delCampo:2017bzr}
\be
\left|\frac{Z(\beta+it)}{Z(\beta)}\right|^2=\prod^N_{k=1}\frac{\cosh(k\omega\beta)-1}{\cosh(k\omega\beta)-\cos(k\omega t)}.
\ee
The harmonic oscillator is the $N=1$ case.

Consequently, the thermal entropy of the CSM can be understood as a summation of thermal entropies of harmonic oscillators with equal spacing frequencies. Therefore, after continuation to imaginary $\beta$, the TPE becomes
\begin{equation}
    S_{N} (\beta + i t; \omega) = \sum_{k=1}^{N} S_{1} (\beta + i t; k \omega).
\end{equation}
On Fig.\,\ref{fig:CSM} (right), we plot the real and imaginary parts for the CS model for $N=2$. The effect of many particles adds more wiggles but the real and imaginary parts keep oscillating periodically with increasing amplitude. The KK relation is again satisfied. 
\subsection{Schwarzian theory}
Our next example, and the first example with continuous spectrum, is the Schwarzian theory. This model has played a prominent role in understanding of the low-energy limit of the Sachdev-Ye-Kitaev (SYK) model \cite{Sachdev:1992fk,kitaev} and its correspondence with Jackiw–Teitelboim gravity \cite{Maldacena:2016hyu} (see also review \cite{Mertens:2022irh}).\\ 
The Schwarzian theory can be defined by the characteristic continuous spectrum with the hyperbolic sine density of states
\be
 \rho_0(E)\equiv\frac{C}{2\pi^2}e^{S_0}\sinh\left(2\pi\sqrt{2CE}\right).
 \label{sinhE}
\ee
In this expression, $C$ and $S_0$ denote large constants and $S_0$ computes the entropy of the 2D black hole. This density of states leads to the Schwarzian thermal partition function that is also one-loop exact \cite{Stanford:2017thb}
\be
Z(\beta)=\int^\infty_0 dE\,\rho_0(E)e^{-\beta E}=\frac{e^{S_0}}{\sqrt{2\pi}}\left(\frac{C}{\beta}\right)^{3/2}e^{\frac{2\pi^2 C}{\beta}}.  \label{SCAq}
\ee
\begin{figure}[t!]
    \centering
    \includegraphics[width=9cm]{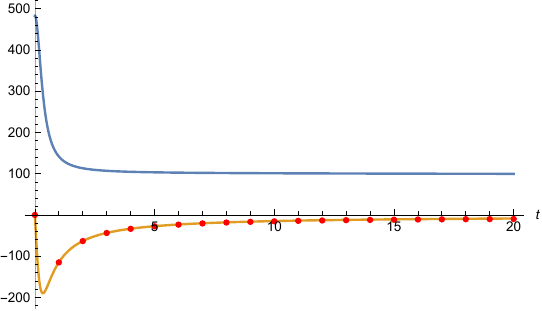}
    \caption{Real (blue) and imaginary (orange) parts of TPE derived from \eqref{SchwPE} for $\beta=1/3$, $S_0=100$ and $C=10$. Red dots indicate numerically computed imaginary part using the KK-relation.}
    \label{fig:SchwarzianTPE}
\end{figure}
Finally, the thermal entropy of the model is given by
\be
S(\beta)=S_0+\frac{4\pi^2 C}{\beta}-\frac{3}{2}\log\left(\beta\right)+\frac{3}{2}\log\left(\frac{C\, e}{(2\pi)^{1/3}}\right),\label{SchwPE}
\ee
and TPE is obtained from it by setting $\beta\to\beta+it$. Fig.\,\ref{fig:SchwarzianTPE} shows its real and imaginary parts as well as the validity check of the KK relation.

From this analytical result we see that, at the early times, TPE is approximately 
\ba
S(\beta+it)\simeq \frac{4\pi^2 C}{\beta+it},
\ea
which can be naturally explained by the classical gravity dual. In the late time regime, we find the logarithmic behaviour 
\ba
S(\beta+it)\simeq -\frac{3}{2}\log t,
\ea
dominated by the one-loop correction.
This behaviour nicely agrees with the general argument (\ref{logpe}). Indeed, for the Schwarzian theory we have the continuous spectrum \eqref{sinhE} which, close to the edge, is characterised by the scaling parameter $\gamma=\frac{1}{2}$.

On the other hand, in the strictly large $t$ limit, we need to respect the discreteness of the spectrum in regular quantum systems. This should be dual to the non-perturbative effects in quantum gravity. Eventually the averaged value of the TPE will approach a certain positive value $S_*$ (\ref{SSW}). 
\subsection{Random matrix theory}
\begin{figure}
    \centering
    \includegraphics[width=7cm]{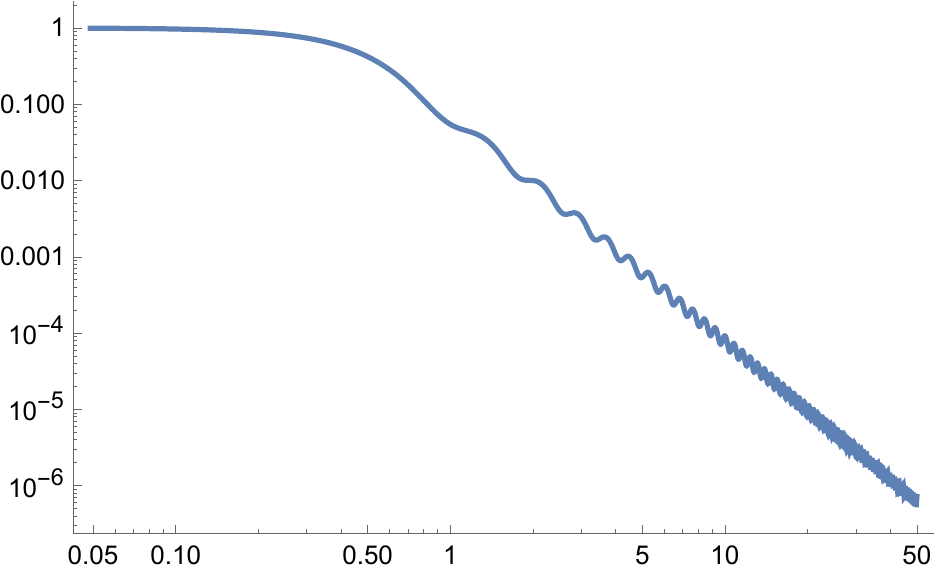}
    \caption{Log-log plot of SFF for RMT with partition function \eqref{ZRMTI1} for $\beta = 0.1$ and $\lambda = 4$.}
    \label{fig:plotRMTSFF}
\end{figure}

Our second example with continuous spectrum will be the random matrix theory (RMT). In general, a RMT Hamiltonian is considered as an archetype of a quantum chaotic setup. We will thus try to draw some universal lessons from this example and compare them with integrable models.\\  More precisely, we consider the quenched partition function from a random ensemble, e.g. GUE. We first check its behaviour in the simplest case of the Wigner semi-circle, leading to a universal logarithmic decay of the real part of the TPE. A numerical analysis shows that taking the discreteness of the spectra, the logarithmic decay stops at the dip time, followed by a logarithmic growth until the late time value $S_{*}$. Later, we discuss why this behaviour is robust against possible non-Gaussian potentials.
\paragraph{Semi-circle}
In RMT, the normalised density of states is usually approximated by the continuous Wigner semicircle spectrum
\begin{equation}
    \rho_{\rm SC}(E) = \frac{2}{\pi \lambda} \sqrt{1 - \frac{E^2}{\lambda^2}},
    \label{semic}
\end{equation}
where $\lambda = \sqrt{2N}$. As a result, the quenched partition function reads
\begin{align}
    Z_{\rm SC}(\beta) = \int^{\lambda}_{-\lambda} dE e^{-\beta E} \rho_{\rm SC}(E)= \frac{2}{\beta \lambda}I_{1}(\beta \lambda).\label{ZRMTI1}
\end{align}
Here, $I_{1}$ is the modified Bessel function of the first kind. It's straightforward to obtain the SFF from the partition function in this approximation
\begin{align}
    {\rm SFF}_{\rm SC}(\beta, t) &= \frac{Z_{\rm SC}(\beta + i t) Z_{\rm SC}(\beta - i t)}{Z_{\rm SC}^{2}(\beta)}.
\end{align}
It is plotted in Fig.\,\ref{fig:plotRMTSFF}, in accordance with the previous results \cite{delCampo:2017bzr}. There is no ramp and plateau (as for finite $N$ RMT \cite{guhr1998random,Cotler:2016fpe}) because the Wigner's semi-circle is derived by first taking the size of the random matrix $N$ to infinity and then normalising with rescaling. Essentially, the gap between energy levels is suppressed in this limit, leading to an infinite dip time.

\begin{figure}[b!]
    \centering
    \includegraphics[width=7cm]{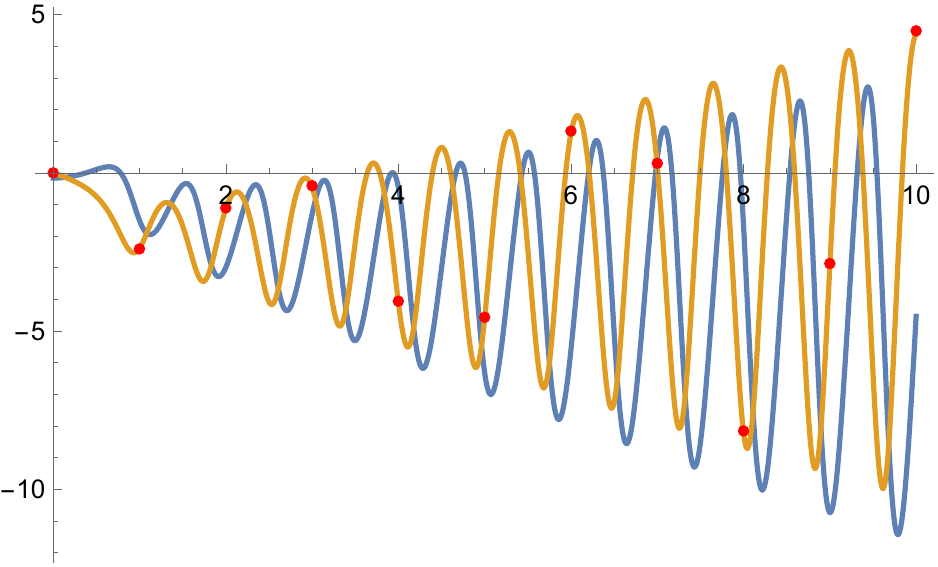}
    \hfill
    \includegraphics[width=7cm]{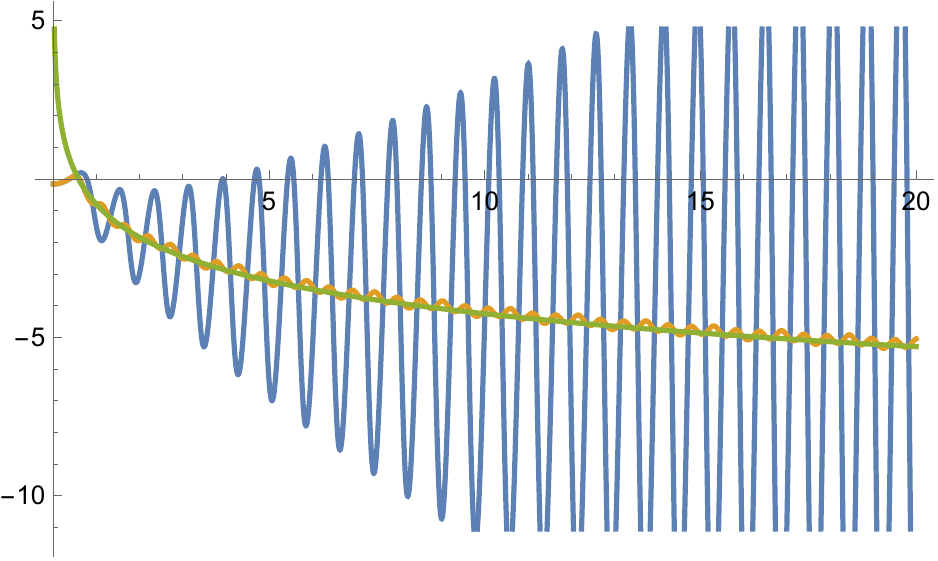}
    \caption{(Left) Re-Im plot of the TPE of random matrix theory with $\beta = 0.3$ and $\lambda = 4$. The KK relation is respected. (Right) Averaged real part of TPE (shown in orange), comparing with the late time prediction \ref{eq:RMTlate} (in green).}
    \label{fig:psRMT}
\end{figure}

\begin{figure}
    \centering
    \includegraphics[width=7cm]{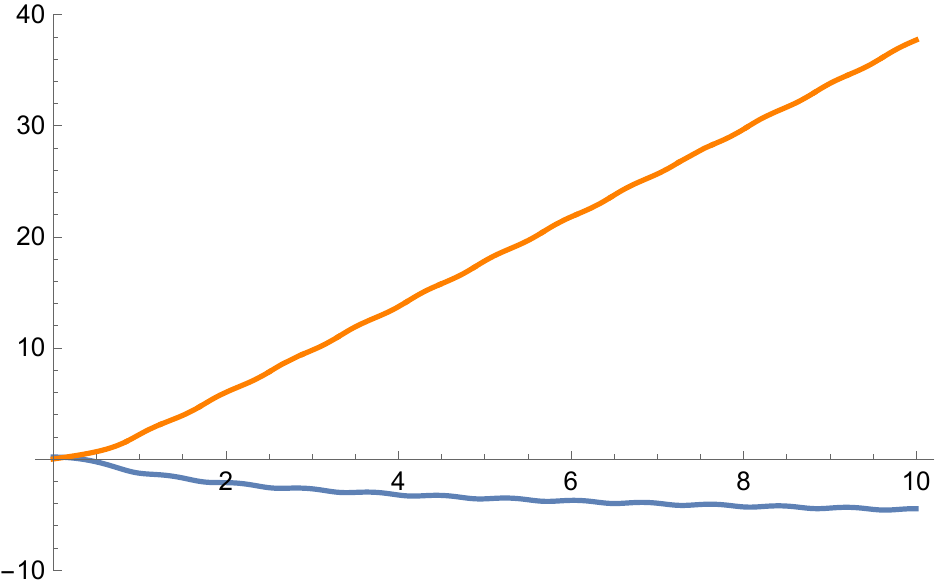}
    \hfill
    \includegraphics[width=7cm]{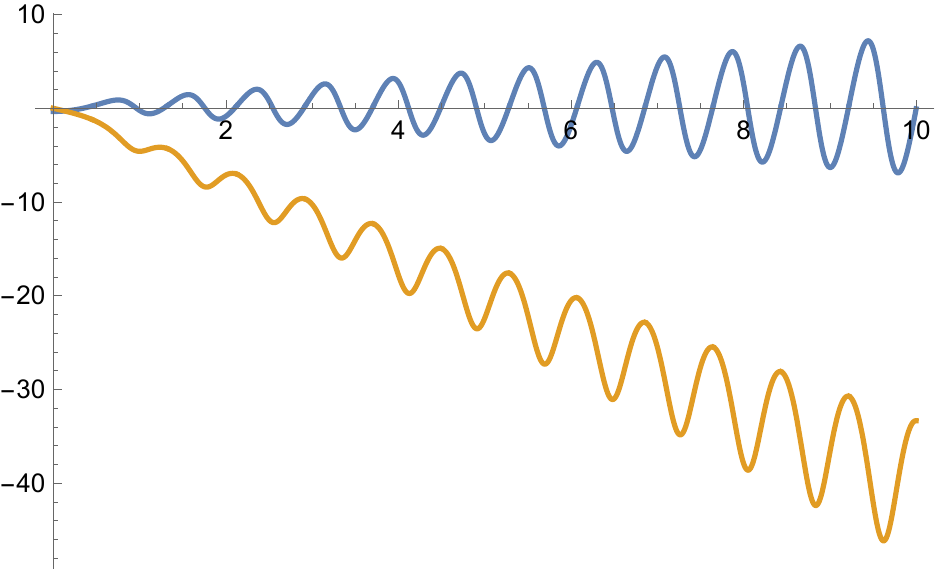}
    \caption{(Left) Re-Im plot of $\log Z$ term for parameters $\beta = 0.3$ and $\lambda = 4$. (Right) Re-Im plot of the energy term.}
    \label{fig:psRMT2}
\end{figure}

Given the partition function, we can obtain the thermal entropy and continue $\beta \rightarrow \beta + it$ to get the TPE. The result is
\begin{equation}
    S_{\rm SC} (\beta + it) = \log \left(  \frac{2}{(\beta + it) \lambda}I_{1}(\left(\beta + it \right) \lambda)  \right) - \lambda (\beta + it) \frac{I_{2}(\left(\beta + it \right) \lambda)}{I_{1}(\left(\beta + it \right) \lambda)},\label{TPERMTI1}
\end{equation}
and we plot its features on Fig.\,\ref{fig:psRMT}.
Both, the imaginary and real parts oscillate with time $t$, with an increasing magnitude. To understand it better, we plot the $\log Z$ term (the first term in \eqref{TPERMTI1}) and the energy term (the second term in \eqref{TPERMTI1}) separately in Fig.\,\ref{fig:psRMT2}. From the plot, it is clear that the imaginary part of the $\log Z$  will accumulate and give a linear growth, which cancels the linear decrease from the energy term.
\\
In fact, this argument can be made more explicit if we consider the large $t$ expansion of TPE analytically. For the modified Bessel function, we have the following large $z$ expansion (${\rm Arg}\, z < \pi/2$)
\begin{equation}
    I_{n}(z) \sim \frac{e^{z}}{\sqrt{2\pi z}} \left( 1 - \frac{4n^{2}-1}{8z} + \dots\right).
\end{equation}
Using this asymptotic formula we can extract the large $t$ behaviour of the TPE as follows
\begin{align}
    S_{\rm SC} ( \beta + i t ) &\sim_{t \gg \beta}  \lambda (\beta + i t) - \frac{3}{2} \log (\beta + i t) + \log \sqrt{\frac{2 }{ \pi \lambda^{3}}} + O((\beta + i t)^{-1}) \notag \\
    &- \lambda (\beta + i t) \left( 1 - \frac{3}{2 \lambda (\beta + i t)} + O((\beta + i t)^{-2})\right)\nonumber \\
    &= - \frac{3}{2} \log (\beta + i t) + \frac{3}{2} + \log \sqrt{\frac{2 }{ \pi \lambda^{3}}} + O((\beta + i t)^{-1}).
    \label{eq:RMTlate}
\end{align}
From the leading order term, we get the scaling with time
\begin{align}
    {\rm Re}[S_{\rm SC} (\beta + i t)] \sim_{t \gg \beta} - \frac{3}{2} \log t, \qquad
    {\rm Im}[S_{\rm SC} (\beta + i t)] \sim_{t \gg \beta} - \frac{3}{4} \pi,
\end{align}
since ${\rm Im} \log (\beta + i t)$ goes to $\pi/2$ and sub-leading terms all go to zero. From Fig.\,\ref{fig:psRMT}, the leading order term of large $t$ expansion (\ref{eq:RMTlate}) resembles the time averaged TPE, which also ignores the details of oscillations. The logarithmic behaviour of the real part again agrees with the general argument in (\ref{logpe}) as the semi circle distribution (\ref{semic}) leads to $\gamma=\frac{1}{2}$. 
It is also straightforward to verify the KK relation in this example
\begin{equation}
    -\frac{3}{2\pi} \int_{-\infty}^{\infty} ds \frac{{\rm Re} [\log t]}{s- t} = -\frac{3}{2\pi} \times 2 \int_{\log t}^{\infty} d u \log \frac{e^u + t}{e^u - t} = -\frac{3 \pi}{4},
\end{equation}
where in the second equality we set $u = \log (s-t)$.
 
\begin{figure}
    \centering
    \includegraphics[width=7cm]{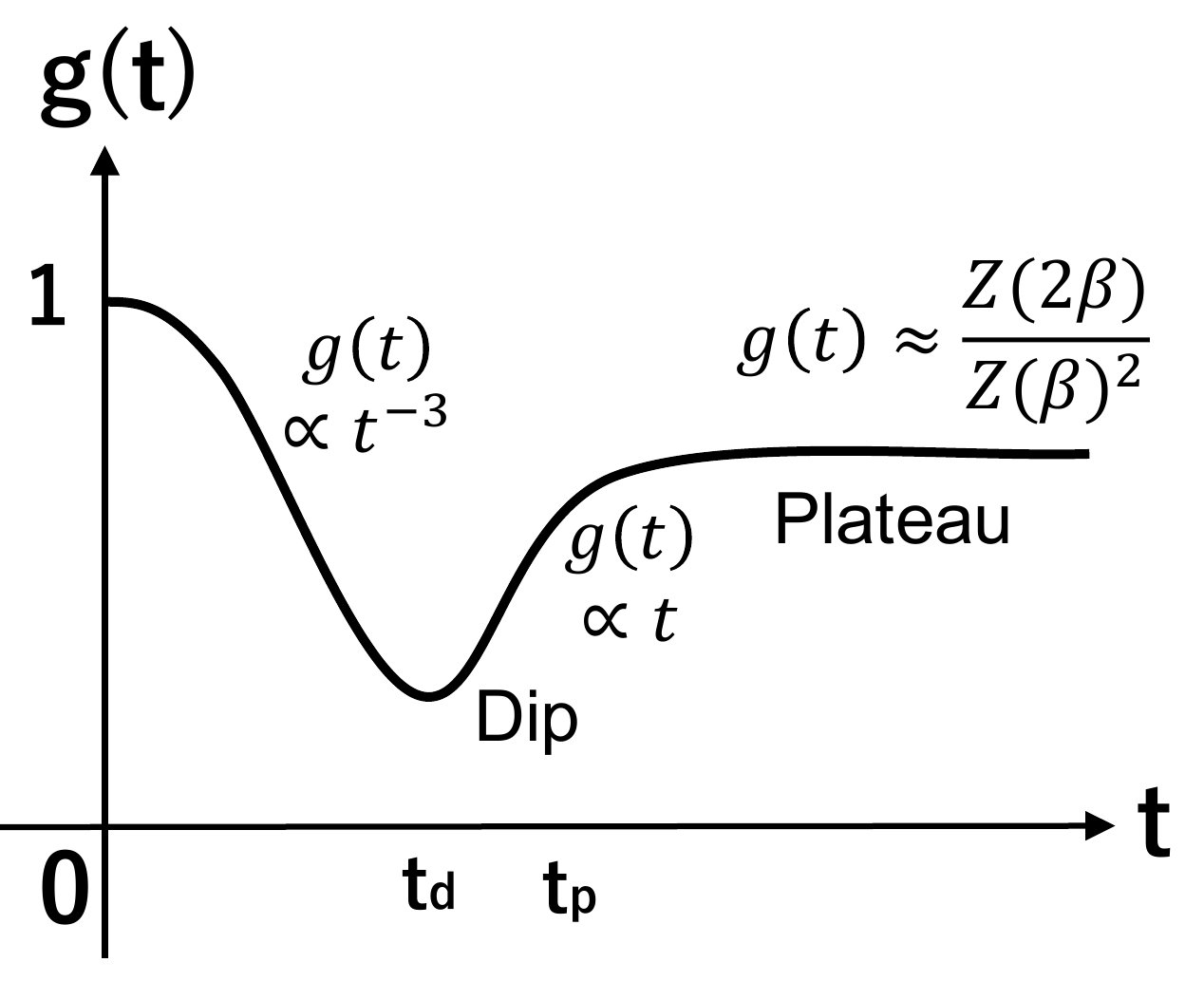}
        \includegraphics[width=8cm]{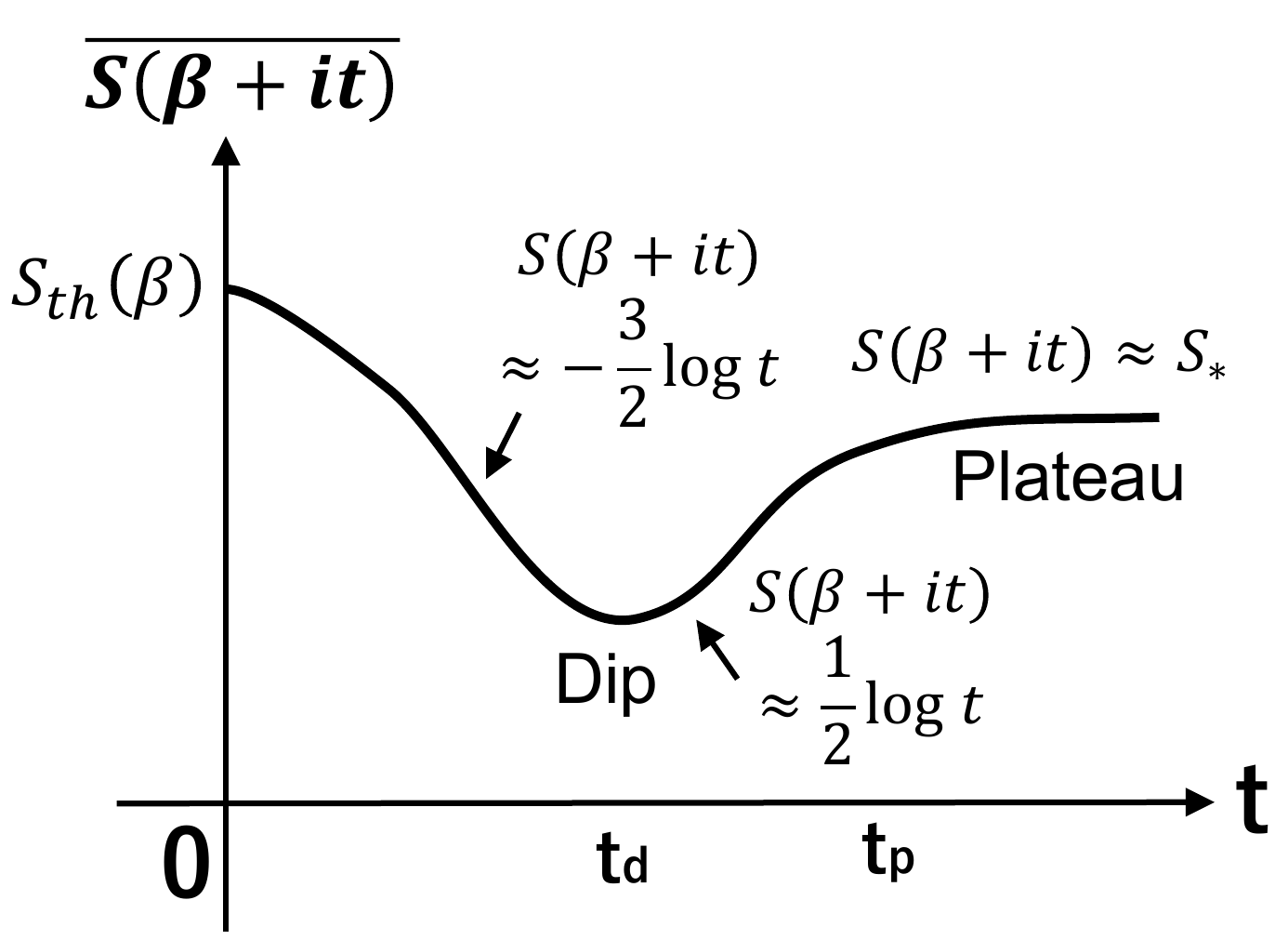}
    \caption{The sketches of evolution of the TPE $g_\beta(t)$ (left) and the averaged TPE $\ov{S(\beta+it)}$ (right) in the random matrix model.}
    \label{fig:TPERMT}
\end{figure}

In the strict large $t$ limit, we will need to take into account the discreteness of the spectrum or non-trivial correlations between eigenvalues. For example, it is well known that in generic quantum chaotic theories, SFF $g_\beta(t)$ will show the dip and ramp behaviour as depicted in the left panel of Fig.\,\ref{fig:TPERMT}. Following the relation (\ref{SFFTPE}), we can use this fact to infer the dynamics of the averaged TPE depicted in the right panel of Fig.\,\ref{fig:TPERMT}. In the ramp region, we expect $g(t)\propto t$ and this leads to $\ov{S(\beta+it)}\simeq \frac{1}{2}\log t$. Finally, the averaged value of the TPE should approach a certain positive value $S_*$ (\ref{SSW}). 

\begin{figure}[h!]
    \centering
    \includegraphics[width=7cm]{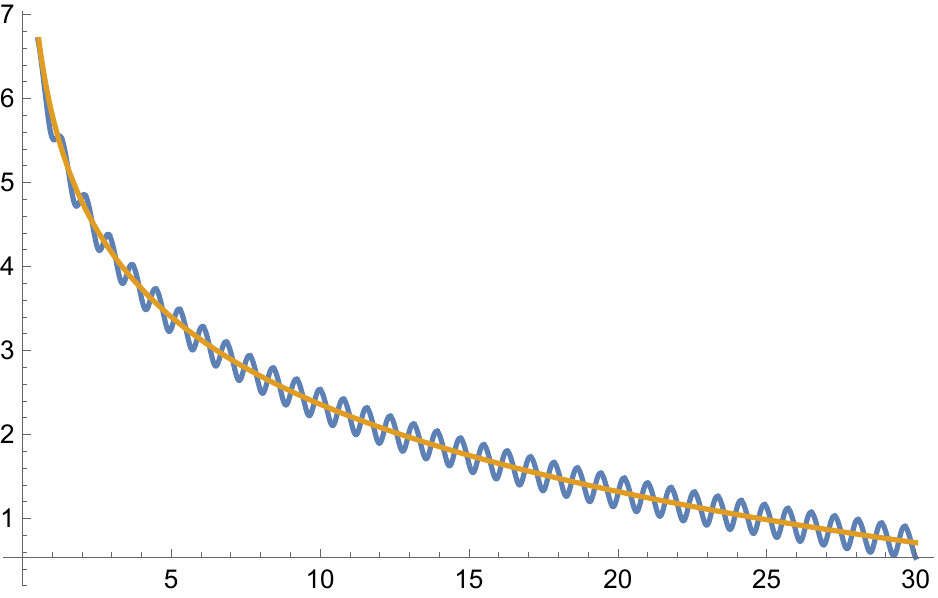}
    \caption{Plot of real parts of the second R\'enyi entropy. The blue curves are exact calculations and the orange curves are late-time answers.}
    \label{fig:RenyiRMT}
\end{figure}

The R\'enyi entropies also show the same features. For simplicity, we consider the second R\'enyi
\begin{equation}
    S^{(2)}_{\rm SC} (\beta + it) = - \log \left[  \frac{Z_{\rm SC}(2 (\beta + it) \lambda)}{Z^{2}_{\rm SC}((\beta + it) \lambda)}    \right] 
    = - \log \left[ \frac{(\beta + it) \lambda}{4} \frac{I_{1}(2 (\beta + it)\lambda)}{I^{2}_{1}((\beta + it)\lambda)} \right],
\end{equation}
that is plotted on Fig.\,\ref{fig:RenyiRMT}. Again, using the asymptotic formula, we can confirm the logarithmic scaling for late time $t$
\begin{equation}
    S^{(2)}_{\rm SC} (\beta + it) \sim -\frac{3}{2} \log (\beta + it) +  \log \frac{4}{\sqrt{\pi} \lambda^{3/2} } + O(1/(\beta + it)).
\end{equation}

\paragraph{Numerics}
To test our conclusions above more thoroughly, every timescale needs to be considered with care. We now present some numerical results of TPE across the whole slope-ramp-plateau region. 

\begin{figure}[b!]
    \centering
    \includegraphics[width=7cm]{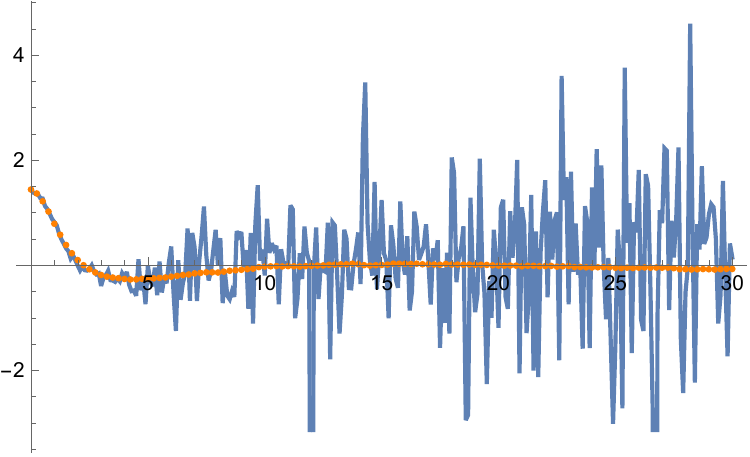}
    \hfill
    \includegraphics[width=7cm]{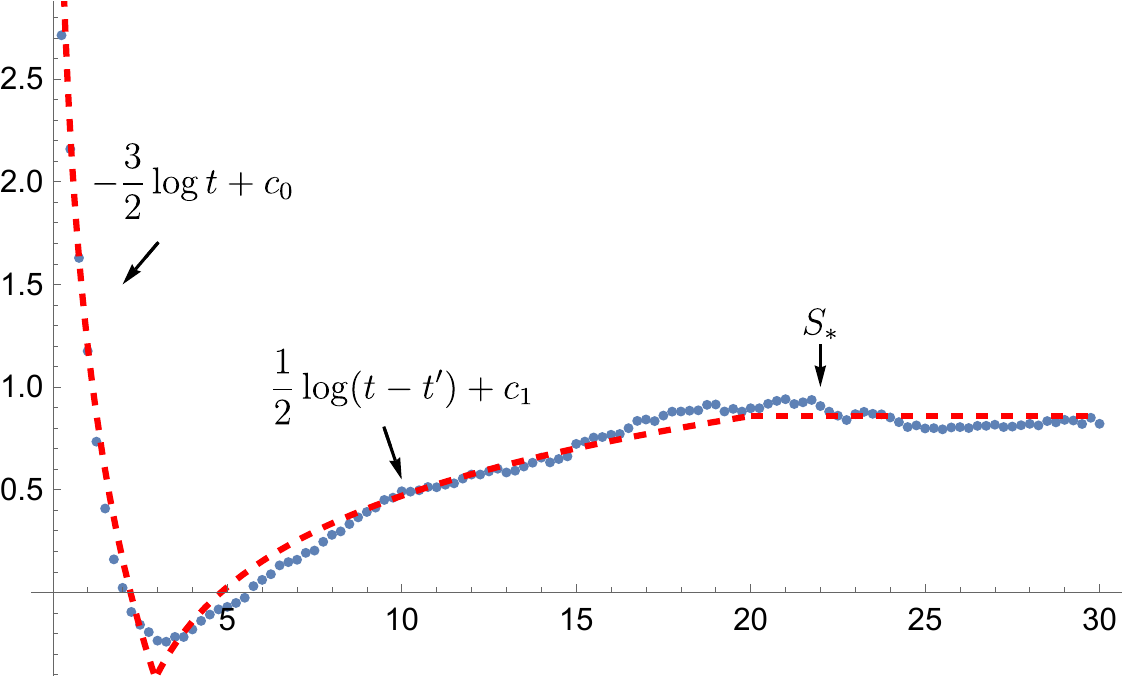}
    \caption{(Left) TPE averages over 200 instances of $100 \times 100$ random unitary matrices. Here $\beta = 1$. A time averaged version is shown in orange. (Right) Time-averaged TPE with the same parameters, fit by the logarithmic behaviours for slope and ramp as well as the final plateau in red dashed lines. }
    \label{fig:RMTnum}
\end{figure}

We take 200 instances of $100 \times 100$ random unitary matrices. First, we compute the partition function of each instance and get the thermal entropy with (\ref{thent}). After a continuation $\beta \rightarrow \beta + i t$, TPE for each instance is found. The TPE is highly oscillatory for a single instance, just like the calculation of SFF. Next, we average over all the instances by taking the mean value of the thermal pseudo-entropies, leading to the left panel of Fig.\,\ref{fig:RMTnum}. In the end, a time-averaging is carried out and is depicted in the right panel of Fig.\,\ref{fig:RMTnum}.

The TPE plot is in general very sporadic, given the increasing oscillating magnitude with $t$. It is the finite-$N$ counterpart of the quenched partition function. After averaging over time, it shows a clear dip-ramp-plateau structure. We sketch the analytical prescription given in Fig.\,\ref{fig:TPERMT}. The slope region is given by $-\frac{3}{2} \log t$ decay, the ramp by $\frac{1}{2} \log (t-t')$ and finally the constant plateau $S_{*}$. The numerical results support the analytical analysis above.
\paragraph{Adding potential}
We can also test the robustness of our analysis by modifying the potential. Namely, the Wigner semi-circle comes from a Gaussian potential $V (M) = \frac{1}{2} M^{2}$. More generically, we can also include higher order terms, e.g. quartic and sextic terms. Here we consider the one-cut sextic case (see \cite{Gaikwad:2017odv} for analysis of the SFF)
\begin{equation}
    V(M) = \frac{1}{2} M^{2} + \frac{g}{N} M^{4} + \frac{h}{N^{2}} M^{6},
\end{equation}
and the density of states is given by
\begin{equation}
    \rho (E) = \frac{1}{\pi} \left( 3 h E^{4} + (2g + 6 h a^{2}) E^{2} + \left(\frac{1}{2} + 4g a^{2} +18 h a^{4} \right)    \right) \sqrt{4 a^{2} - E^{2}}.
\end{equation}
The normalisation gives the following constraint on the couplings
\begin{equation}
    60 h a^{6} + 12 g a^{4} + a^{2} - 1 =0.
\end{equation}
For example, $g=h=0$ and $a=1$ give us the semi-circle solution.\\
The partition function for this model becomes
\bea
    Z_{\rm sextic}(\beta) &=& \left( \frac{360 h a^{3}}{\beta^{3}} + \frac{a + 24 g a^{3} + 180 h a^{5}}{\beta} \right) I_{1}(2a \beta)\nonumber\\ 
    &-& \left( \frac{720 h a^{2}}{\beta^{4}} + \frac{24 g a^{2} + 360 h a^{4}}{\beta^{2}} \right) I_{2}(2a \beta).
\ea
Following the same steps as for the Gaussian model, we can obtain the TPE and its large time scaling 
\begin{align}
    S_{\rm sextic}(z) &\sim \log \left[ e^{2 a z} \left( \frac{\sqrt{a} (1 + 24 g a^{2} + 180 h a^{4})  }{2\sqrt{\pi} z^{3/2}}  - \frac{ 3 (1 + 152 g a^{2} + 2100 h a^{4}) }{32 \sqrt{a \pi} z^{5/2} } + \dots \right) \right]\nonumber \\
    &-  2 a z + \frac{3}{2} - \frac{3(1 + 152 g a^{2} + 2100 h a^{4})}{16(a + 24 g a^{3} +  180 h a^{5}) z} + \dots  \nonumber \\
    &= - \frac{3}{2} \log z + {\rm const.} + O(1/z),
\end{align}
where in the end we just replace $z$ by $\beta + i t$.

This confirms that adding higher order terms in the potential does not modify the leading order behaviour of TPE. The $-\frac{3}{2} \log t$ is preserved under perturbations. As is expected from the discussion in (\ref{logpe}), the logarithmic behaviour only depends on the scaling of spectrum close to the ground state edge. RMT is equivalent to $\gamma = \frac{1}{2}$, regardless of the details of the potential. The same can be shown for multi-cut RMT spectra and the universal logarithmic behaviour is still preserved.
\subsection{Averaged thermal pseudo-entropy for the RMT universality class}
In this subsection we revisit the above claims, but this time more quantitatively and analytically in the RMT universality class.\\
We saw that the average value of the real part of TPE is directly related to the SFF via
\be
\overline{\text{Re}\,S^{(n)}(\beta+it)}=\frac{\log\langle Z(n(\beta+it))Z(n(\beta-it))\rangle-n\log\langle Z(\beta+it)Z(\beta-it)\rangle}{2(1-n)}.\label{ExprAvSFF}
\ee
In this expression we assumed that we can take the average under the logarithm (no replica symmetry breaking). We will proceed with this reasonable assumption that holds in many physical systems\footnote{It will also be important to study TPE without this assumption, e.g. \cite{Derrida:1981zz}. See, related \cite{Winer:2022ciz}.}.\\
In RMT we can take the advantage of \eqref{ExprAvSFF} to predict the evolution of $\overline{\text{Re}\,S^{(n)}}$ at all times. Indeed, this can be derived in complete analogy to the SFF, pedagogically reviewed in  \cite{Mertens:2022irh}, that we closely follow with slight modifications for our context\footnote{Analysis of TPE in analogy with \cite{Winer:2020gdp} would also be interesting. }.

The basic object that we will compute is
\be
\langle Z(n(\beta+it))Z(n(\beta-it))\rangle=\int dEdE'\langle\rho(E)\rho(E')\rangle e^{-n\beta(E+E')}e^{-int(E-E')}.
\ee
The RMT universality posits that the two-point correlator of densities $\rho(E)$ is universal (irrespectively of the potential) in the unitary universality class, and given by
\be
\langle \rho(E)\rho(E')\rangle\simeq \rho_0(E)\rho_0(E')-\frac{\sin^2\left(\pi\rho_0(E)(E-E')\right)}{\pi^2(E-E')^2}+\rho_0(E)\delta(E-E'),
\ee
where the density $\rho_0(E)$ is identical to the Schwarzian theory \eqref{sinhE}. The second term is the sine kernel responsible for the level repulsion in quantum chaotic systems. We will separate it using the identity $\sin^2(x)=(1-\cos(2x))/2$ as
\be
\langle \rho(E)\rho(E')\rangle\simeq \rho_0(E)\rho_0(E')-\frac{1}{2\pi^2(E-E')^2}+\frac{\cos\left(2\pi\rho_0(E)(E-E')\right)}{2\pi^2(E-E')^2}+\rho_0(E)\delta(E-E'),\label{DensFin}
\ee
and analyse each of the contributions separately.\\
For early times, the product of two densities dominates. This gives the product of two Schwarzian partition functions such that
\be
\langle Z(n(\beta+it))Z(n(\beta-it))\rangle\simeq\frac{C^3e^{2S_0}}{2\pi n^3(t^2+\beta^2)^{3/2}}e^{\frac{4\pi^2 C\beta}{n(t^2+\beta^2)}},
\ee
and the real part of the averaged n-th R\'enyi TPE becomes
\be
\overline{\text{Re}\,S^{(n)}(\beta+it)}\simeq S_0+\frac{n+1}{n}\frac{2\pi^2C\beta}{t^2+\beta^2}-\frac{3}{4}\log(t^2+\beta^2)+\frac{3}{2}\frac{\log(n)}{n-1}+\frac{3}{2}\log\left(\frac{C}{(2\pi)^{1/3}}\right).
\ee
After the collision time $t\sim \beta$ the SFF decays as $t^{-3}$, what translates to the logarithmic decay of the TPE. Note that the $-3/2\log(t)$ contribution is the same for all the R\'enyi entropies i.e. independent of $n$.\\
Next we turn to the second term in \eqref{DensFin} for which we have
\be
\langle Z(n(\beta+it))Z(n(\beta-it))\rangle\simeq-\int dEdE' \frac{1}{2\pi^2(E-E')^2} e^{-n\beta(E+E')}e^{-int(E-E')}.
\ee
This integral is also universal\footnote{We first change integration variables to $(E,E-E')$, integrate over $E$ and use the Fourier transform identity $-\int^{+\infty}_{-\infty}\frac{dx}{2\pi^2x^2}e^{iTx}=|T|/2\pi$ for $T=-n(t+i\beta)$.} and independent on $n$. It gives \cite{Ginsparg:1993is,Mertens:2022irh}
\be
\langle Z(n(\beta+it))Z(n(\beta-it))\rangle\simeq\frac{\sqrt{t^2+\beta^2}}{4\pi\beta}\to\frac{t}{4\pi\beta}.
\ee
Interestingly, on the gravity side, this term is associated with the Lorentzian double cone geometry \cite{Saad:2018bqo,Chen:2023hra} which also fixes the normalisation. Consequently, the TPE changes evolution from the logarithmic decay to the dip to the logarithmic ramp regime (see Fig.\,\ref{fig:TPERMT}) 
\be
\overline{\text{Re}\,S^{(n)}(\beta+it)}\simeq\frac{1}{2}\log\left(\frac{t}{4\pi\beta}\right).
\ee
Again we get the universal $1/2\log(t)$ growth (together with the normalisation) for all the R\'enyis. However, the dip time is estimated from the exchange of dominance between the two contributions
\be
\frac{C^3e^{2S_0}}{2\pi n^3t^3_d}=\frac{t_d}{4\pi\beta}\quad\Rightarrow\quad t_d= (2\beta C^3)^{1/4}e^{S_0/2}/n^{3/4},
\ee
and depends on the R\'enyi index $n$.\\
The third term in \eqref{DensFin} is non-perturbative in $e^{-S_0}$ and, for the SFF, results in a linear decay that cancels the linear ramp contribution \cite{Mertens:2022irh}. It is estimated for times of order
\be
t\sim Ce^{S_0},
\ee
and this is the plateau time. Finally, the value of the plateau is obtained from the last term
\be
\langle Z(n(\beta+it))Z(n(\beta-it))\rangle\simeq\int dE\rho_0(E) e^{-2n\beta E}=Z(2n\beta),
\ee
and corresponds to the TPE plateau exactly as in \eqref{LWW}.\\

This reproduces our claims from the previous sections and Fig.\,\ref{fig:TPERMT} in the RMT universality class.
\section{Thermal pseudo-entropy in  two-dimensional CFTs}\label{sec:2dCFTs}
In this section, we begin the analysis of TPE in a larger class of examples, namely in two-dimensional conformal field theories (CFT). In particular, we focus on two extreme families: the free and the holographic 2D CFTs. We will separately discuss the symmetric product orbifold CFTs, whose properties somehow sit in the middle between the two, in the next section. Since the SFF have not been analysed in all of these models before, we also present them when needed.
\subsection{Free Fermion CFT}\label{sec:2dCFTsf}
\begin{figure}[t!]
    \centering
    \includegraphics[width=5cm]{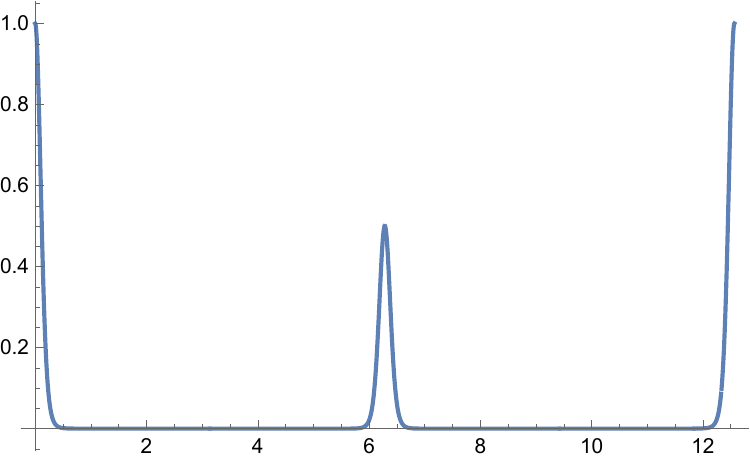}
    \includegraphics[width=5cm]{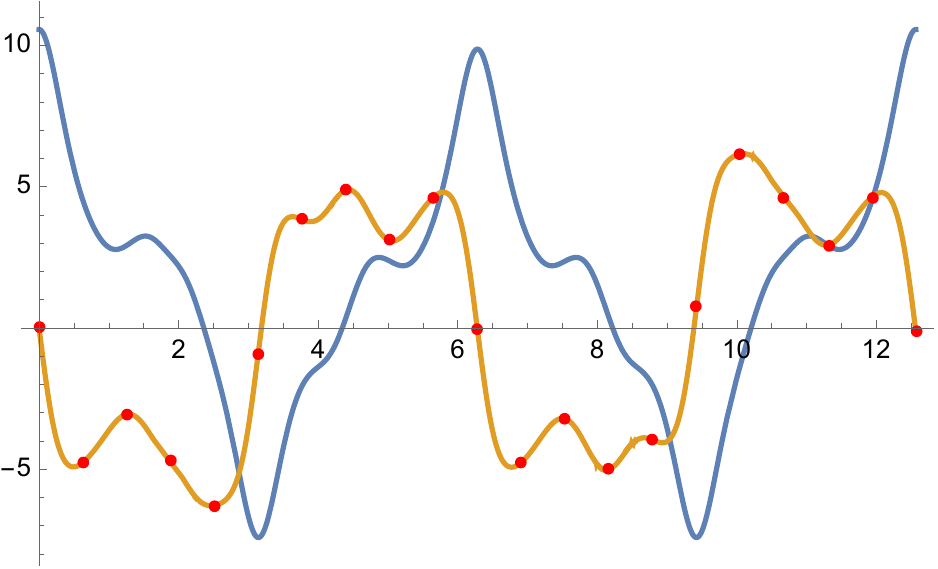}
    \vspace{5mm}
    \includegraphics[width=5cm]{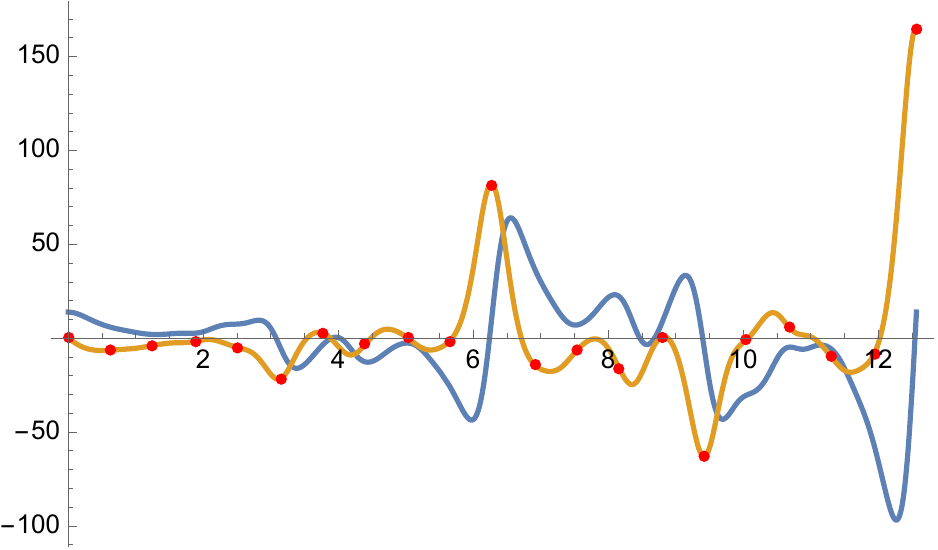}
    \caption{The plots of time evolution of the SFF $g_{\beta}(t)$ (left), the second R\'enyi TPE $S^{(2)}(\beta+it)$ (middle), and the (von-Neumann) TPE $S(\beta+it)$ (right)
    for the $c=1$ Dirac fermion CFT. The blue and orange curves correspond to the real and imaginary parts of the entropies, respectively. Red dots confirm the KK-relations. Plots for $\beta=1/2$.}
    \label{fig:DiracC=1}
\end{figure}
In the $c=1$ Dirac fermion CFT or free boson CFT, the modular invariant partition function depends on the complex modular parameter $\tau$ and reads \cite{Yellow}
\be
Z(\tau,\bar{\tau})=\frac{|\theta_3(\tau)|^2+|\theta_2(\tau)|^2+|\theta_4(\tau)|^2}{|\eta(\tau)|^2}.
\ee
Correspondingly, the thermal partition function at temperature $T=1/\beta$ is given by 
\be
Z_{th}(\beta)=Z\left(\frac{i\beta}{2\pi},-\frac{i\beta}{2\pi}\right).
\ee
In Fig.\,\ref{fig:DiracC=1}, we plotted of the SFF
$g_\beta(t)$, the second R\'enyi TPE $S^{(2)}(\beta+it)$, and the TPE $S(\beta+it)$. Note that the first two follows the exact periodicity $t\sim t+4\pi$, while the amplitude of the final one (TPE) grows linearly.
In this example, we do not find the logarithmic behaviour (\ref{logS}) because there is a finite energy gap between the vacuum and the first excited state.
\subsection{Compact scalar CFT}
\label{sec:FreeScalarCFT}
Next, we consider a free compact scalar $\phi$ with the compactification radius $R$, i.e. we impose the identification $\phi\sim\phi+2\pi R$. The partition function reads \cite{francesco2012conformal}
\ba
Z_{R}(\tau.\bar{\tau})=\frac{1}{|\eta(\tau)|^2}\sum_{n,w\in Z}
e^{2\pi i\tau\left(\frac{n}{2R}+\frac{wR}{2}\right)^2}
e^{-2\pi i\bar{\tau}\left(\frac{n}{2R}+\frac{wR}{2}\right)^2}, \label{patsc}
\ea
where we set $\tau=\frac{i\beta}{2\pi}$ with the inverse temperature $\beta$. 

\begin{figure}[h!]
    \centering
    \includegraphics[width=3.7cm]{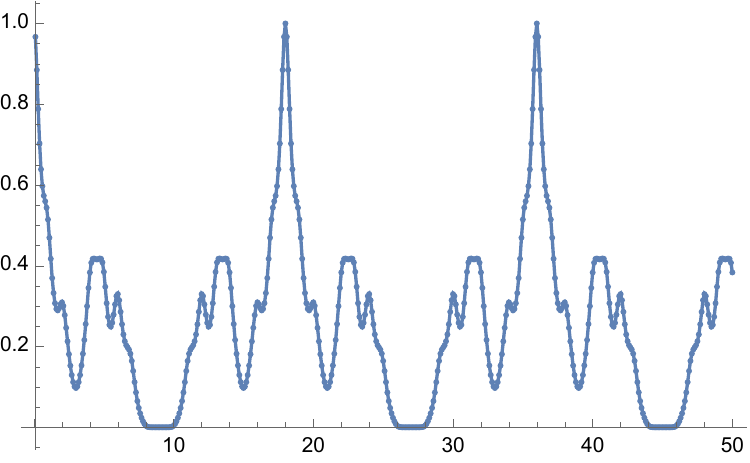}
    \hfill
    \includegraphics[width=3.7cm]{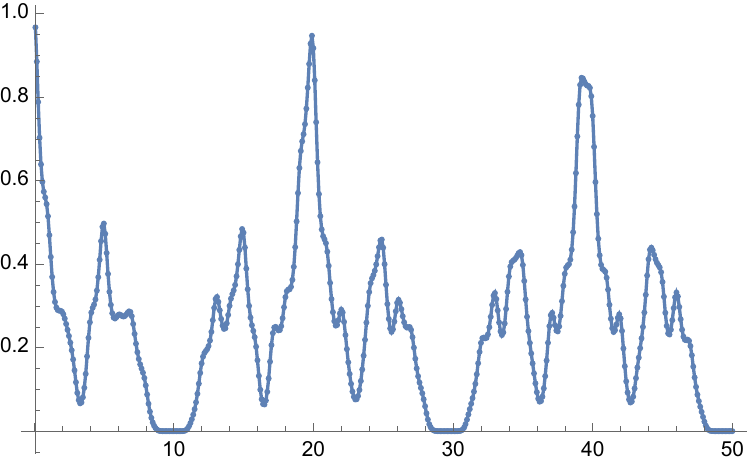}
    \hfill
    \includegraphics[width=3.7cm]{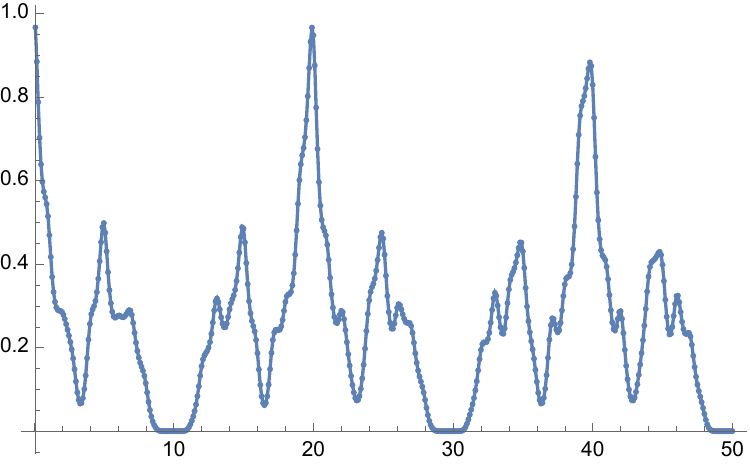}
    \hfill
    \includegraphics[width=3.7cm]{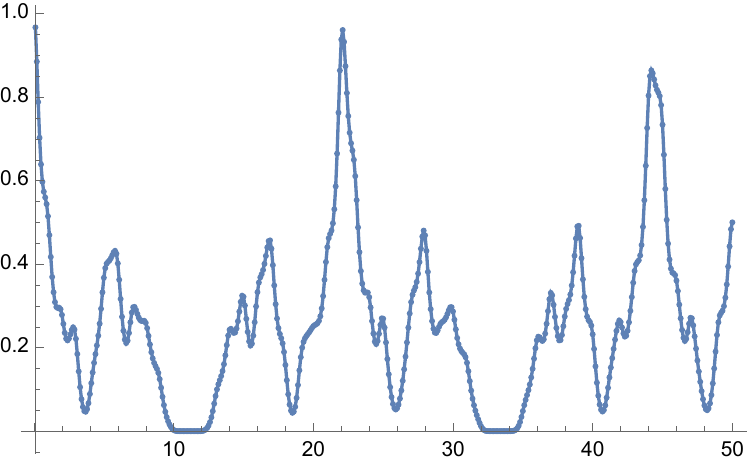}
    \caption{SFFs for compact scalar CFTs as a function of $\frac{t}{2\pi}$. From the left: $R$ is $3$, $\pi$, $\sqrt{2}+\sqrt{3}\sim 3.14626437$, $0.3$, respectively. Plots for the inverse temperature $\beta/2\pi=0.6$.}
    \label{fig:CompScSFF}
\end{figure}

\begin{figure}[h!]
    \centering
    \includegraphics[width=5cm]{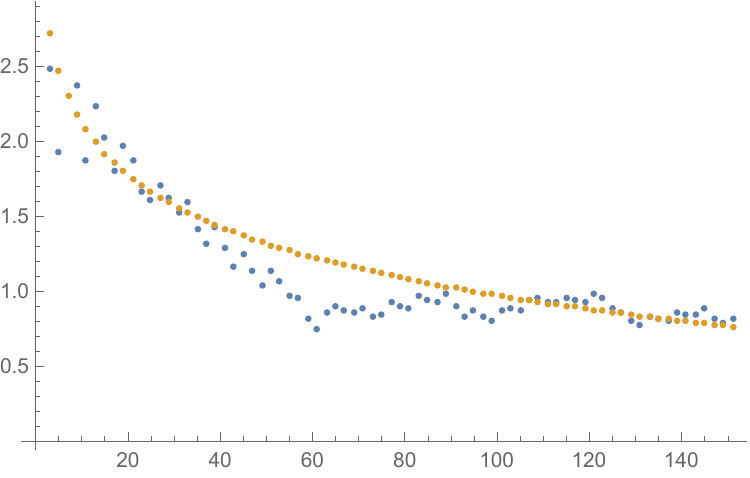}
    \hfill
    \includegraphics[width=5cm]{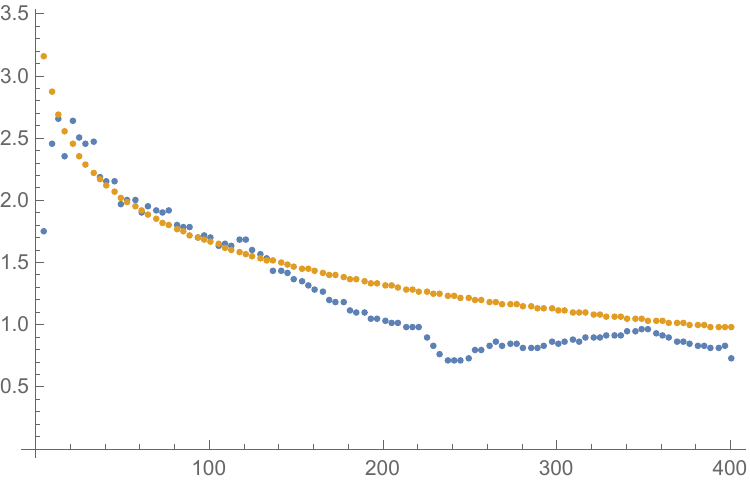}
    \hfill
    \includegraphics[width=5cm]{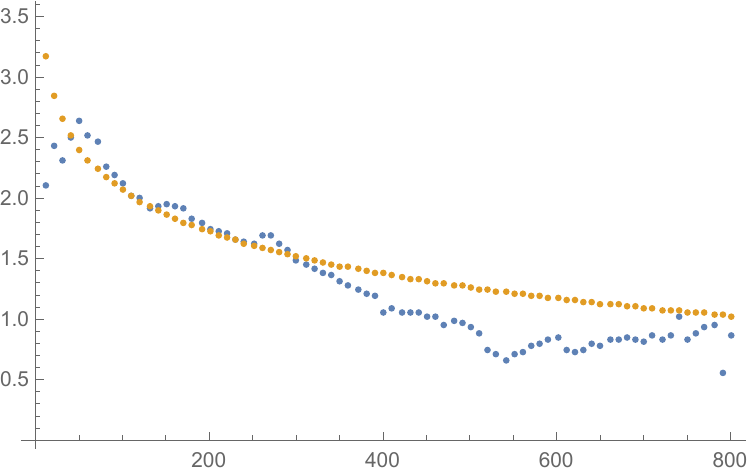}
    \hfill
    \caption{Plots of time averaged  second R\'enyi TPE for the compact free scalar CFT at $\beta=0.3$. We chose the compactification radius to be $R=10$ (left), $R=20$ (middle) and $R=30$ (right). The blue dots show the plot of numerical values and the orange curve is a fit with the 
    $-\frac{1}{2}\log t+$ const. For each values of $R$, the fit matches nicely up to $t=25, 100$ and $225$, respectively.} \label{fig:TPEscp}
\end{figure}

The SFF for this CFT can be evaluated numerically as shown in Fig.\,\ref{fig:CompScSFF}, where, $n, w$ in (\ref{patsc}) are restricted to $-100 \leq n,w \leq 100$ for this numerical calculation.  They are almost periodic, and their period $T$ can be determined analytically: $\frac{T}{2\pi} = \max\{2R^2,2/R^2\}$. As we can see from the figures, the early time behaviour does not depend on the rationality of $R$.
\\
In the decompactification limit $R\to \infty$, the partition function becomes
\ba
Z_{R=\infty}(\beta)=\frac{V_1}{(2\pi\beta)^{\frac{1}{2}}}\frac{1}{|\eta(\tau)|^2},
\ea
where $V_1=\lim_{R\to\infty}2\pi R$. In this non-compact scalar case, we find $\gamma=-\frac{1}{2}$. Similarly, in case of $N$ non-compact scalars we would have the following late time behaviour for the TPE
\ba
\ov{S(\beta+it)}\simeq -\frac{N}{2}\log t.
\ea
For the compact scalar with a large, but finite, radius $R$ we expect to get the above $\log t$ behaviour up to the time $t\ll \frac{1}{\Delta E}\sim R^2$. This can be confirmed from the plots in 
Fig.\,\ref{fig:TPEscp}, where the time averaged value of second R\'enyi TPE 
\begin{align}
  \overline{S^{(2)}}\left(\beta+it\right) = \frac{1}{t}\int^t_0 dt' ~ S^{(2)}\left(\beta+it'\right),
    \label{tavepe}
\end{align}
was numerically evaluated.
\subsection{Holographic CFTs}
\label{sec:HoloCFT}
Now we turn to analyse the opposite limit i.e. the holographic CFT with a large central charge and in the strong coupling limit. Since we can only treat this class of theories analytically in the classical gravity limit, our analysis becomes qualitative.

In general, for a holographic CFT dual to pure gravity in AdS$_3$, we expect the following spectrum 
\ba
\rho(E)=\rho_0(E)+\rho_{ex}(E),
\ea
where $\rho_0(E)$ represents the contributions from the vacuum state and its descendants. On the other hand, $\rho_{ex}(E)$ describes the excited states contributions with $E\geq \frac{c}{12}$, which behaves according to the Cardy formula
\ba
\rho_{ex}(E)\sim e^{2\pi\s{\frac{c}{3}(E-\frac{c}{12})} }.
\ea
Although we can also relax the behaviour of density of states so that the theory includes primaries with a sparse spectrum for $E<\frac{c}{12}$ \cite{Hartman:2014oaa,Mukhametzhanov:2019pzy}, the conclusions below would not change. 

Then the holographic CFT partition function looks like
\ba
Z(\beta)=Z_0(\beta)+Z_{ex}(\beta),
\label{first}
\ea
where $Z_0(\beta)$ describes the vacuum sector contributions dual to the thermal AdS$_3$. Its dependence on $\beta$ is 
\be
Z_0(\beta)=|q|^{-\frac{c}{24}}\frac{|1-q|^2}{|\eta(\tau)|^2}\simeq e^{\frac{c}{12}\beta},
\ee
where $q=e^{2\pi i\tau}$ and $\tau=\frac{i}{2\pi}$. In the final expression we took advantage of the large $c$ limit of the holographic CFT.

On the other hand, $Z_{ex}(\beta)$ describes the contribution to the partition function from excited states. In the high temperature phase $\beta<2\pi$, it is dominated by the BTZ black hole saddle $Z_{BTZ}(\beta)$ \cite{Banados:1992wn}, obtained by the modular S-transformation $\tau\to -\frac{1}{\tau}$ in the thermal AdS partition function:
\ba
Z_{BTZ}(\beta)= Z_0\left(\frac{4\pi^2}{\beta}\right)\simeq e^{\frac{\pi^2 c}{3\beta}},
\ea
Note that there is no power of $\beta$ in front of $e^{\frac{\pi^2 c}{3\beta}}$ and thus there is no logarithmic term (i.e. $\log \beta$) in the thermal entropy, given by $S_{th}(\beta)=\frac{2\pi^2 c}{3\beta}$, as opposed to the result in the Schwarzian theory.

As usual, we are assuming the black hole phase $\beta<2\pi$ at $t=0$. In the early time region, $t\ll 1$, we can then easily compute the TPE as follows
\ba
S(\beta+it)=\frac{2\pi^2 c}{3(\beta+it)}.
\ea
However, for evaluating the TPE in the late time region $t>1$, the black hole saddle approximation breaks down and other saddle points, including the thermal AdS, start to contribute. Moreover, we do not expect $\log t$ behaviour as we noted just before.\\ 
In other words, we have $\gamma=-1$ in the holographic CFTs. This shows a sharp contrast with the standard ``quantum chaotic" theories i.e. RMT or the Schwarzian, where we found $\gamma=1/2$. 

This way, the full analysis of SFF as well as the TPE in holographic CFTs turns out to be very difficult. This motivates us to study a simpler but partially similar class of two-dimensional CFTs, namely the symmetric orbifold CFTs, as we will do in the next section. Even though the symmetric orbifolds are still free from interactions, they share the property of the sparse spectrum with the actual holographic CFTs. Also the recurrence time becomes exponentially large \cite{TaTsu}. We will explore their imprints in the SFF and TPE below.
\section{Symmetric Orbifold CFTs}\label{sec:Orbifold}
In this section, we turn to the orbifolds of the free CFTs discussed in the previous section. In particular, we compute the SFF and TPE in the symmetric orbifold CFTs. These CFTs are obtained by taking symmetric orbifold projections on $N$ copies of the original seed theory
\begin{equation}
    \mathcal{C}_{N,S} = \mathcal{C}^{\otimes N}/S_{N}.
\end{equation}
In this way, we can construct a large-$N$ CFT from a simple CFT with a small central charge, e.g. a free fermion or a free boson. Despite the fact that large-$N$, symmetric orbifold CFT has a Hagedorn transition, it is not holographically dual to Einstein gravity in asymptotically AdS$_3$ spacetimes.  Indeed, it is not chaotic due to the lack of interactions between different copies, as shown by the calculations of the Lyapunov exponent \cite{Perlmutter:2016pkf} and the evolution of entanglement entropy \cite{Caputa:2017tju}. Nevertheless, recently large-$N$ symmetric orbifold CFTs have turned out to have string theory duals \cite{Gaberdiel:2018rqv,Eberhardt:2018ouy,Eberhardt:2019ywk}. Thus, it is interesting to test how TPE evolves for these complicated, yet integrable theories.

Given the partition function of a seed CFT $Z (\tau , \bar{\tau})$, the partition function of its symmetric orbifold reads
\begin{equation}
    Z_{N,S} (\tau , \bar{\tau} ) = \sum_{\{ N_{k} \}} \prod_{k=1}^{N} \frac{1}{N_{k} !} \left( T_{k} Z( \tau , \bar{\tau})   \right)^{N_{k}},
\end{equation}
where $\{N_{k}\}$ constitutes a partition of integer $N$ so that $\sum_{k} k N_{k} = N$ and $T_{k}$ is the Hecke operator which action on the partition function is defined as
\begin{equation}
    T_{k} Z( \tau , \bar{\tau}) \equiv \frac{1}{k} \sum_{i|k} \sum_{j=0}^{i-1} Z\left( \frac{k \tau}{i^2} + \frac{j}{i} , \frac{k \bar{\tau}}{i^2} + \frac{j}{i}\right).
\end{equation}
For example, when $N=2$ we can write down the orbifold partition function as
\bea
    Z_{2,S} ( \tau , \bar{\tau}) &=& \frac{1}{2!} \left(T_{1} Z( \tau , \bar{\tau})  \right)^{2} + T_{2} Z( \tau , \bar{\tau})\nonumber \\
    &=& \frac{1}{2} \left[ Z^{2}( \tau , \bar{\tau}) + Z( 2 \tau , 2\bar{\tau}) +  Z\left( \frac{\tau}{2} , \frac{\bar{\tau}}{2}\right) + Z\left(\frac{\tau +1 }{2} , \frac{\bar{\tau} + 1}{2}\right) \right].\label{N2SymOr}
\eea

Next we will derive the partition functions of symmetric orbifolds for different $N$ and evaluate thermal pseudo-entropies. The seed theories we will choose are free fermion CFT and compact scalar CFT, the same as in the previous sections.
\subsection{Free Fermion CFT}\label{sec:freefermion}
We first analyse the SFFs for symmetric orbifold CFTs defined as 
\begin{equation}
    {\rm SFF}_{N,S}(\beta, t) = \frac{Z_{N,S} \left(\frac{i (\beta +i t)}{2 \pi},-\frac{i (\beta +i t)}{2 \pi}\right) Z_{N,S} \left(\frac{i (\beta -i t)}{2 \pi},-\frac{i (\beta -i t)}{2 \pi}\right)}{Z^{2}_{N,S} \left(\frac{i \beta }{2 \pi},-\frac{i \beta }{2 \pi}\right) }.
\end{equation}
The SFF is periodic in $t$, similarly to what we already saw on Fig.\,\ref{fig:DiracC=1}. However, as $t$ increases, the log-log plot of the SFF becomes sporadic. Therefore, it is useful to consider a time average of the quantity, as is previously considered in \cite{Balasubramanian:2016ids}.
\begin{figure}[t!]
    \centering
    \includegraphics[width=10cm]{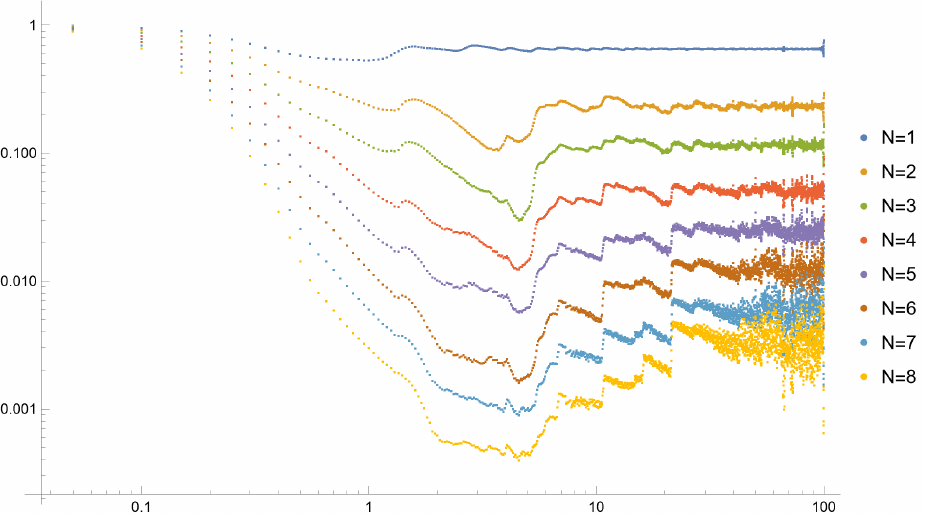}
    \caption{Averaged SFF of free-fermion symmetric orbifolds with increasing $N$. Here $\beta = 0.6$.}
    \label{fig:symorbsffN}
\end{figure}

The log-log plot of the progressive time averaged SFFs from $N=1$ to $N=8$ is shown in Fig.\,\ref{fig:symorbsffN}, with $\beta = 0.6$ and $t \in (0,100)$. We normalise the SFF respectively so that they all start from 1 for $t=0$. Here are some common features
\begin{itemize}
    \item With increasing $N$, the value of the late-time plateau is decreasing. This is simply because the plateau value $Z(2\beta)/Z(\beta)^2$ decreases with $N$ monotonically.
    \item Different from the RMT, there are more steps (also called bumps in \cite{Chen:2022hbi}) in the ramp region. As $N$ increases, there are more steps showing up.
    \item The ``dip'' can be seen as the first step. We also observe that the positions of steps are identical.
\end{itemize}
\begin{figure}[h!]
    \centering
    \includegraphics[width=8cm]{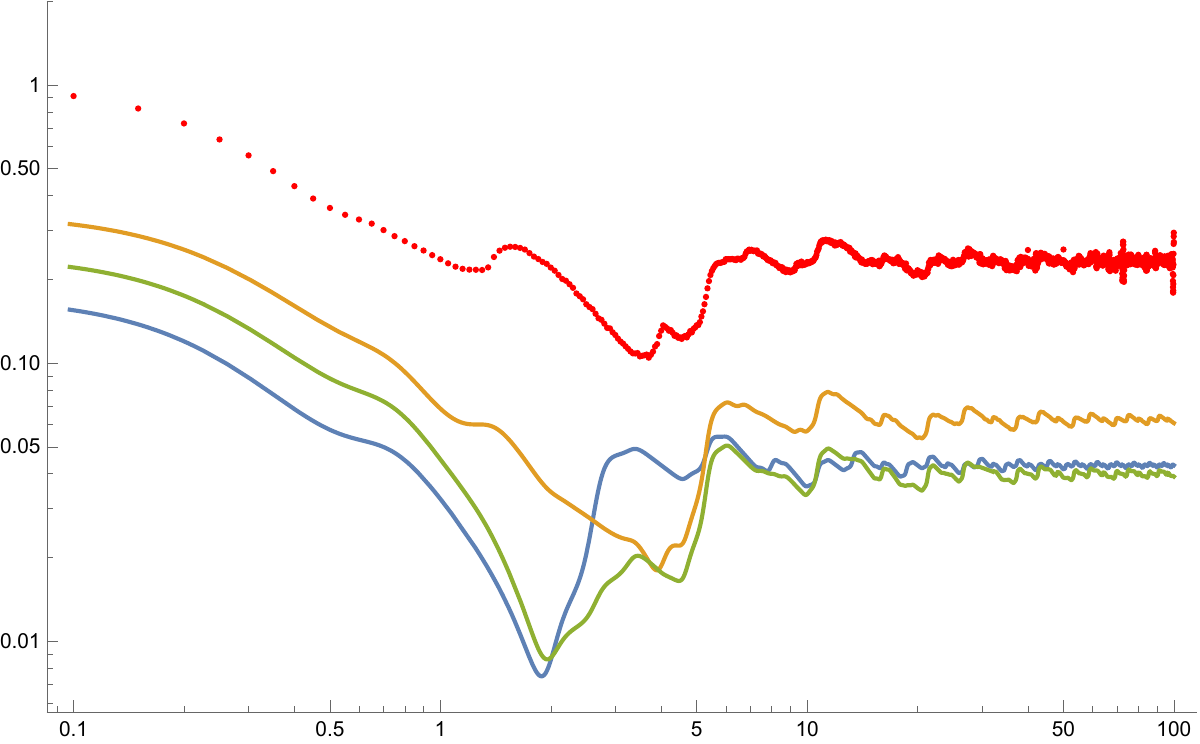}
    \caption{Time evolution of the SFF and contributions from different terms \eqref{N2SymOr} for the $N=2$ orbifold CFT example. Here $\beta = 0.6$.}
    \label{fig:dips}
\end{figure}

A tentative explanation of the above features is shown in Fig.\,\ref{fig:dips}. For simplicity, we choose the $N=2$ case and plot its SFF in red. Considering the partition function of the symmetric orbifold CFT is a sum of terms with increasing Hecke operators, we plot the terms of $T_{1}Z$ in blue, $T_{2}Z$ in orange, and the cross terms in green. We find that the step is to balance the ``slope'' and ``ramp'' from different sectors.

For each term, these behaviours are govern by the presence of peaks before the averaging and a fixed term, which is a polynomial of $T_{i}Z$s in general, is included in any SFF for a large enough $N$. This explains why there are more steps as we increase $N$, since more sectors are included. Lower sectors are always included in the higher sum, which indicates the positions of the steps are almost invariant. \cite{Chen:2022hbi} also discussed the steps in the ramp regime for string gases. It was argued that the positions of steps are related to different winding numbers of the Euclidean Horowitz-Polchinski solutions. It would be interesting to understand our results for symmetric orbifolds along this way too and we leave it for the future.

\begin{figure}[t!]
    \centering
    \includegraphics[width=7cm]{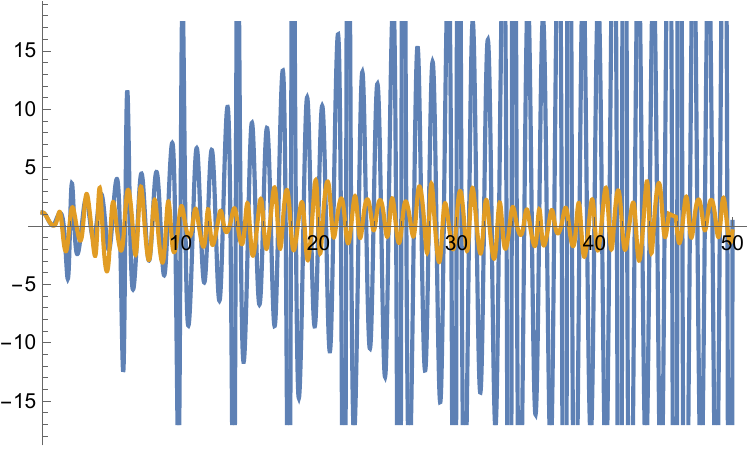}
    \hfill
    \includegraphics[width=7cm]{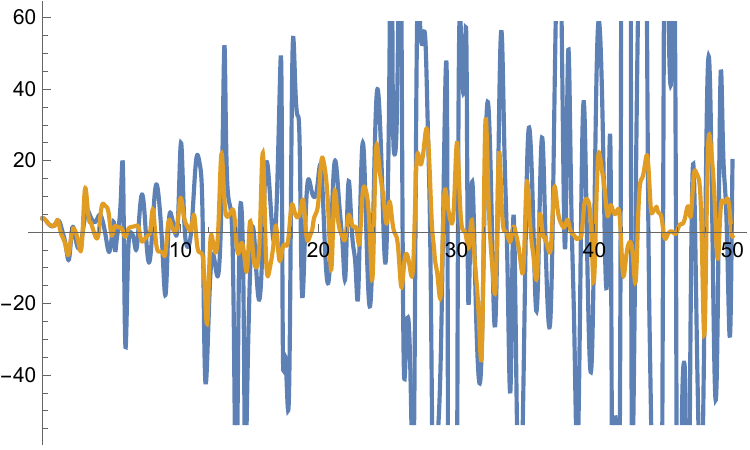}
    \caption{Plots of real parts of TPE for $N=1$ (left) and $N=3$ (right) for $\beta = 0.6$, with the time averaging shown in orange.}
    \label{fig:tpeorbf}
\end{figure}
Next, we evaluate and plot the real part of TPE in Fig.\,\ref{fig:tpeorbf}. Similarly, the magnitude of oscillations increases with time. However, the time averaging does not show any logarithmic decay. This is because already the seed fermion CFT has a large spectrum gap, as we have seen in section \ref{sec:2dCFTsf}.

\subsection{Compact scalar CFT}\label{sec:SFF}
\begin{figure}[h!]
    \centering
    \includegraphics[width=6cm]{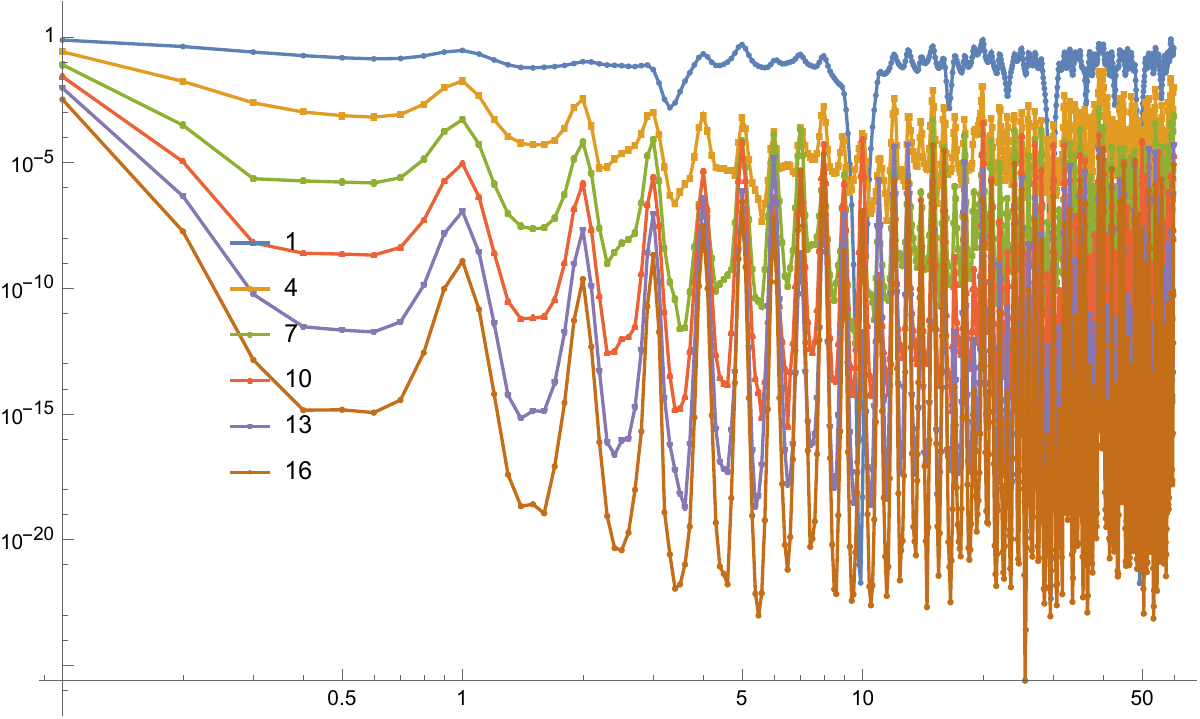}
    \qquad
    \includegraphics[width=6cm]{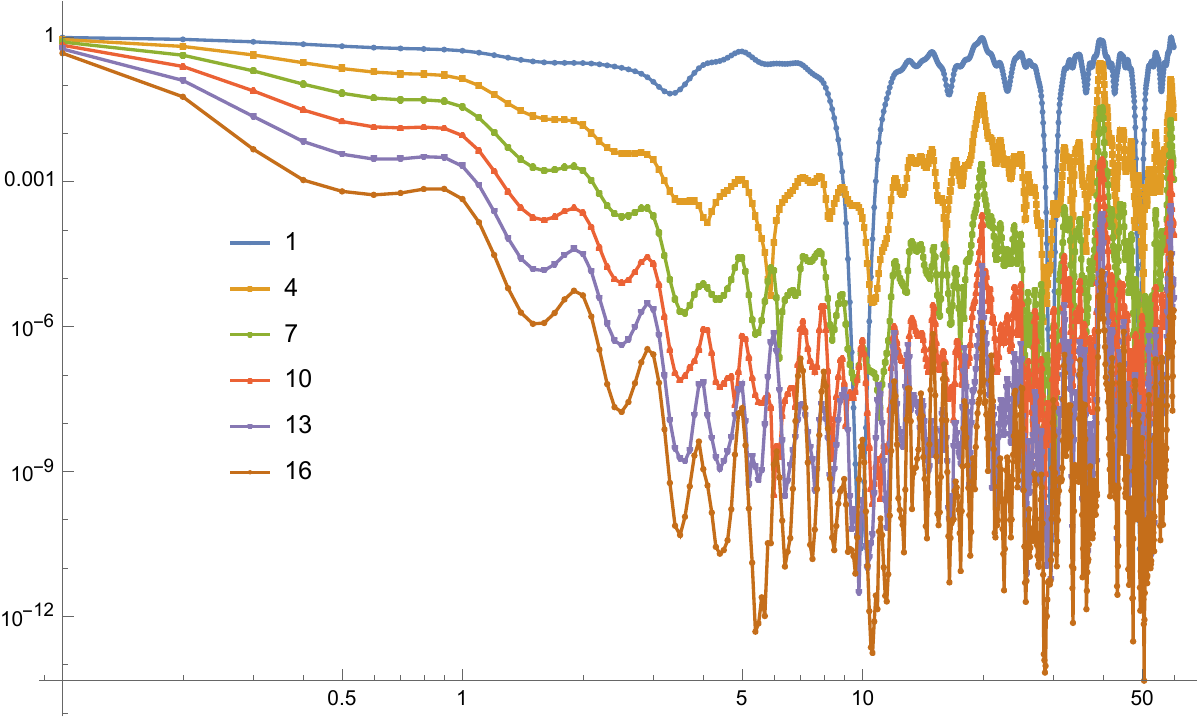}
        \\\vspace{3mm}
    \includegraphics[width=6cm]{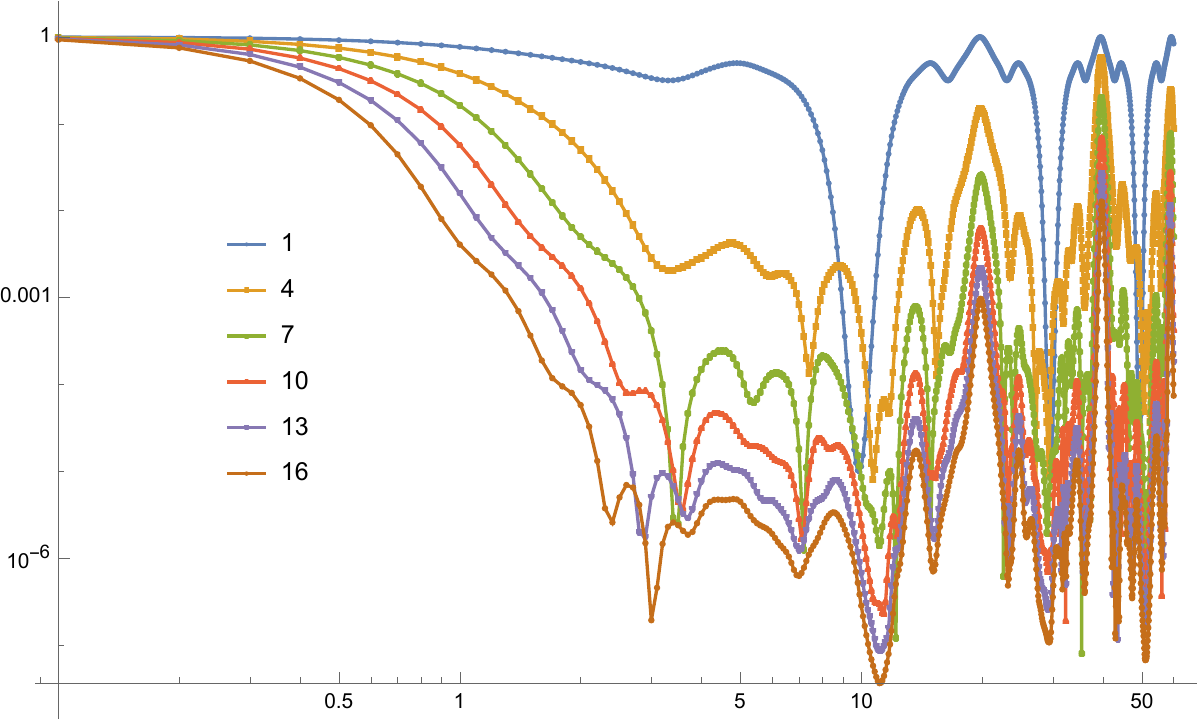}
   \qquad
    \includegraphics[width=6cm]{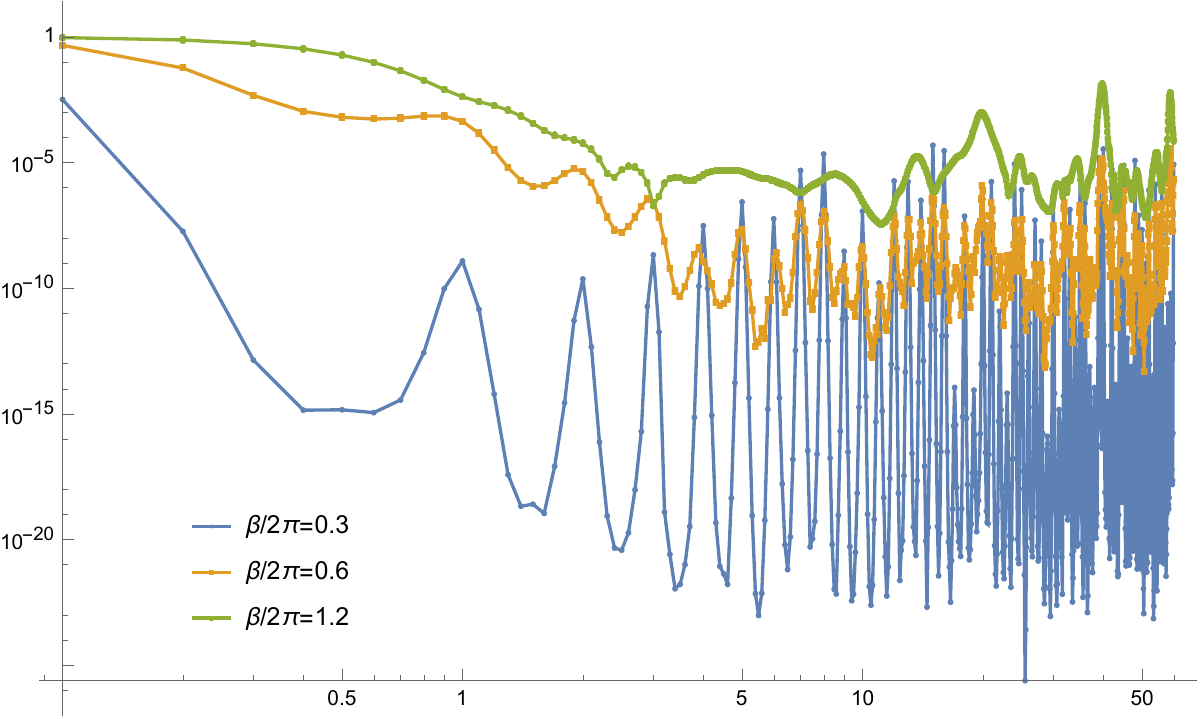}
    \caption{SFFs of compact scalar symmetric orbifolds as a function of $\frac{t}{2\pi}$. $R$ is set to be $\pi$. Upper left figure is for $\frac{\beta}{2\pi} = 0.3$, and right for $0.6$. Lower left for $\frac{\beta}{2\pi} =1.2$, with increasing $N$ and lower right is for $N=16$ with different $\beta$.}
    \label{fig:CompScSymSFF1}
\end{figure}

\begin{figure}[t!]
    \centering
    \includegraphics[width=6cm]{SymOrbPlots/SFF_RisPi_Bis0.6_Nis1-16_noAver.pdf}
    \qquad
    \includegraphics[width=6cm]{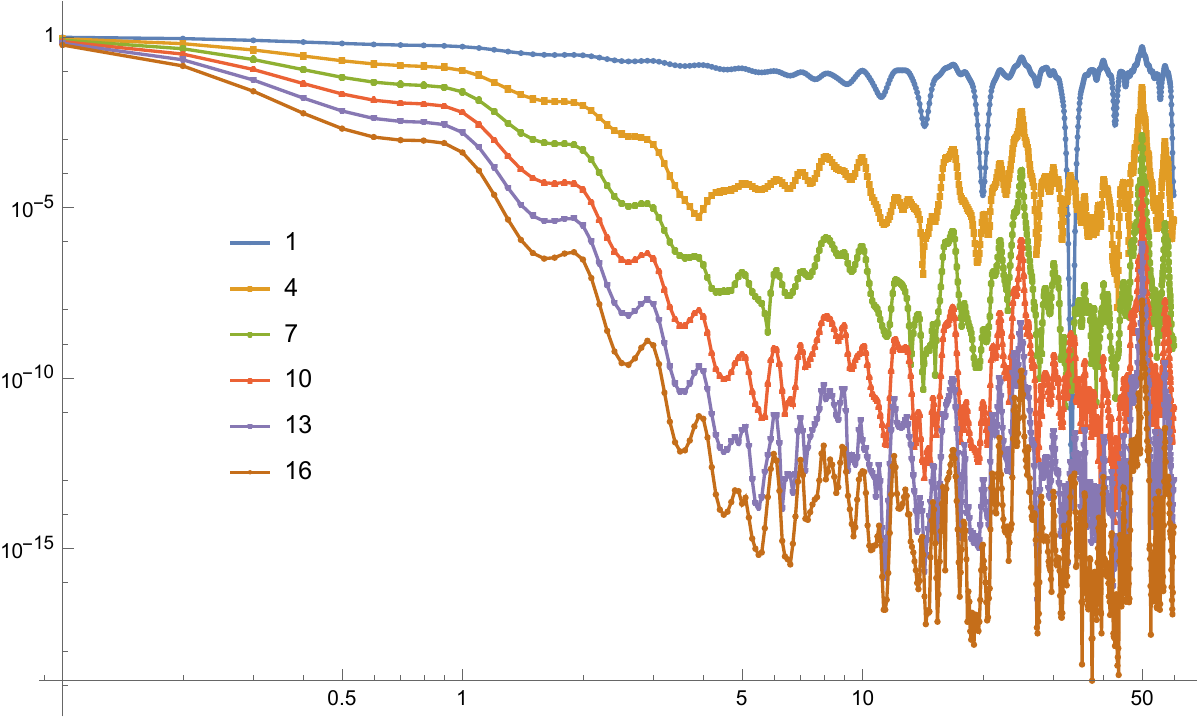}
    \\\vspace{3mm}
    \includegraphics[width=6cm]{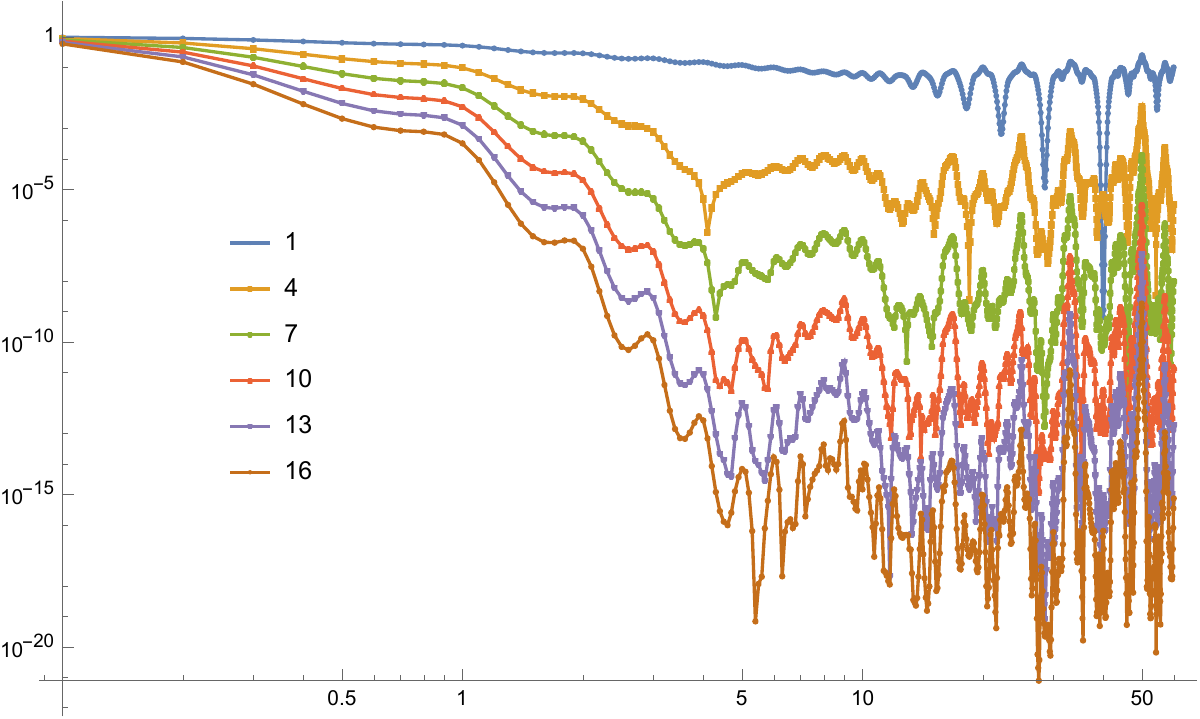}
    \qquad
    \includegraphics[width=6cm]{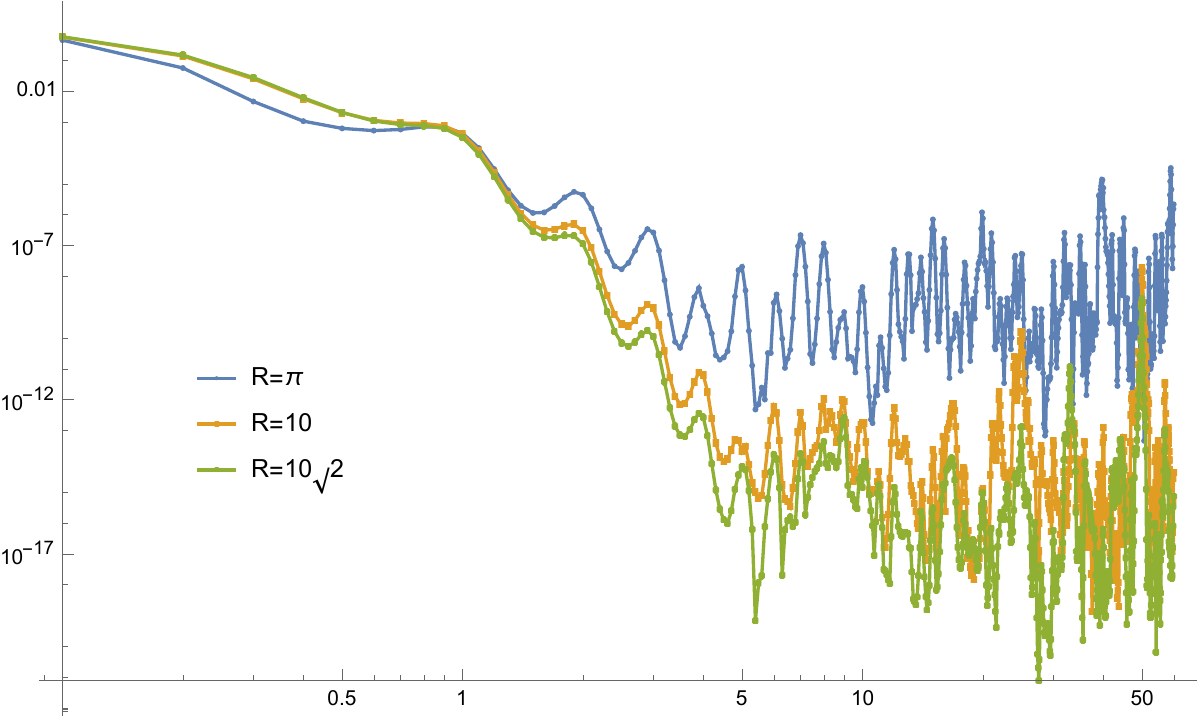}
    \caption{SFFs of compact scalar symmetric orbifolds as a function of $\frac{t}{2\pi}$. $\frac{\beta}{2\pi}$ is set to be $0.6$. Upper left figure is for $R = \pi$, right for $R = 10$. Lower left for $R = 10\sqrt{2}$, with increasing $N$ and right for $N=16$ with different $R$.}
    \label{fig:CompScSymSFF2}
\end{figure}

Now let us move on to the analysis of compact scalar symmetric orbifolds. 
In this section we use the same seed partition function $Z_R(\tau,\bar{\tau})$ as in \eqref{patsc}. For the calculations below, $n, w$ are again restricted to $-100 \leq n,w \leq 100$ when we numerically evaluate SFF and TPE. A complete list of figures is given in the Appendix \ref{sec:CompactScalarSymOrbFig}. Here, we only choose the plots for three different values of the compactification radius $R=\pi,10$ and $10\s{2}$. The log-log plots of the SFFs are shown in Fig.\,\ref{fig:CompScSymSFF1} and Fig.\,\ref{fig:CompScSymSFF2}, with $\frac{t}{2\pi} \in (0,60)$. 

\begin{figure}[b!]
    \centering
    \includegraphics[width=8cm]{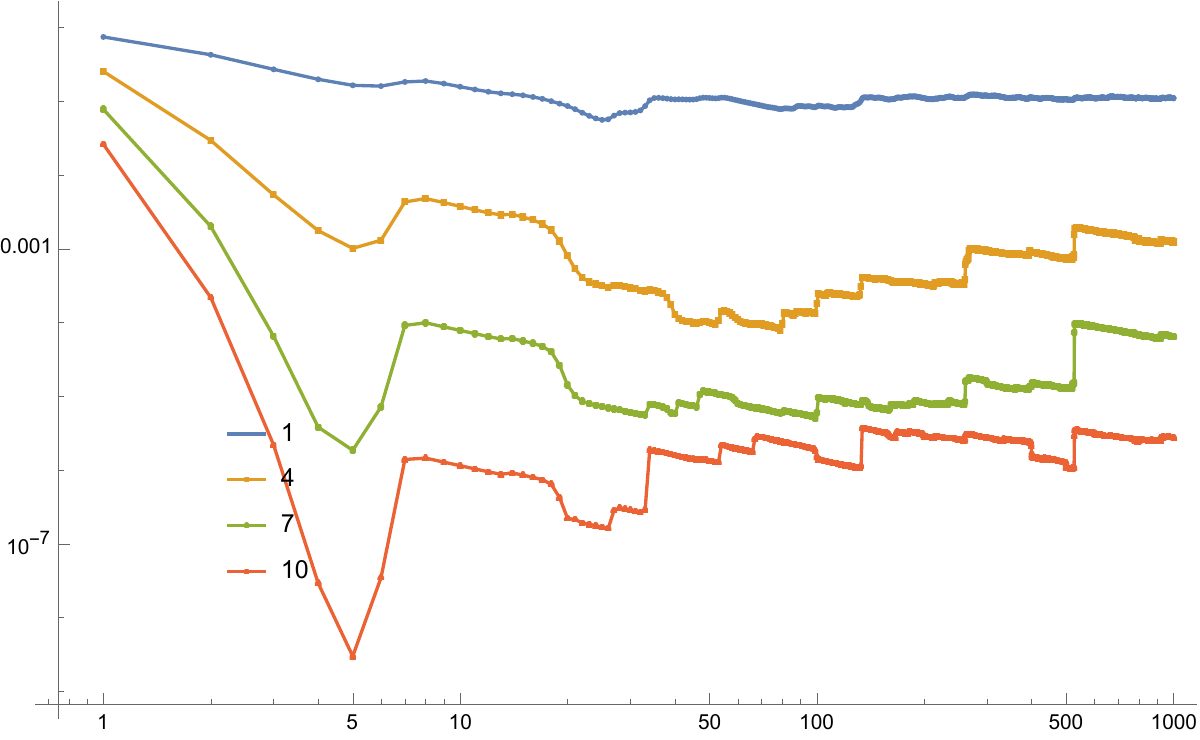}
    \caption{Averaged SFF of compact scalar symmetric orbifolds with increasing N, as a function of $\frac{t}{2\pi}$. Here $R=\pi$ and $\frac{\beta}{2\pi} = 0.3$. For technical reasons, the time scale is 10 times larger than it should be.}
    \label{fig:SFFCSaver}
\end{figure}

Examples of $N$ that we plot are set to values of 1, 4, 7, 10, 13, and 16. The log-log plot of the progressive time averaged SFFs for $N=1,4,7,10$ is shown in Fig.\,\ref{fig:SFFCSaver}, with $R=\pi$, $\frac{\beta}{2\pi} = 0.3$ and $\frac{t}{2\pi} \in (0,100)$. SFFs are normalised as well.
\\

The oscillations become stronger for higher temperatures. Additionally, the value of $t$ that gives the peak of the oscillation is universal i.e. independent of the value of $R$, as is shown in Fig.\,\ref{fig:CompScSymSFF2}.
This suggests that the main contribution to SFF in large-$N$ symmetric orbifolds is from the twist sector. 
Fig.\,\ref{fig:SFFCSaver}, the averaged SFF of compact scalar symmetric orbifolds, shares the same features with Fig.\,\ref{fig:symorbsffN}, the averaged SFF of free-fermion symmetric orbifolds. We can observe the ``dip'' and ``ramp" in spite of the heavy oscillating behaviour. This fact confirms our expectation that SFF of symmetric orbifolds are seed-independent in some sense, especially in large $N$.

\begin{figure}[h!]
    \centering
    \includegraphics[width=5cm]{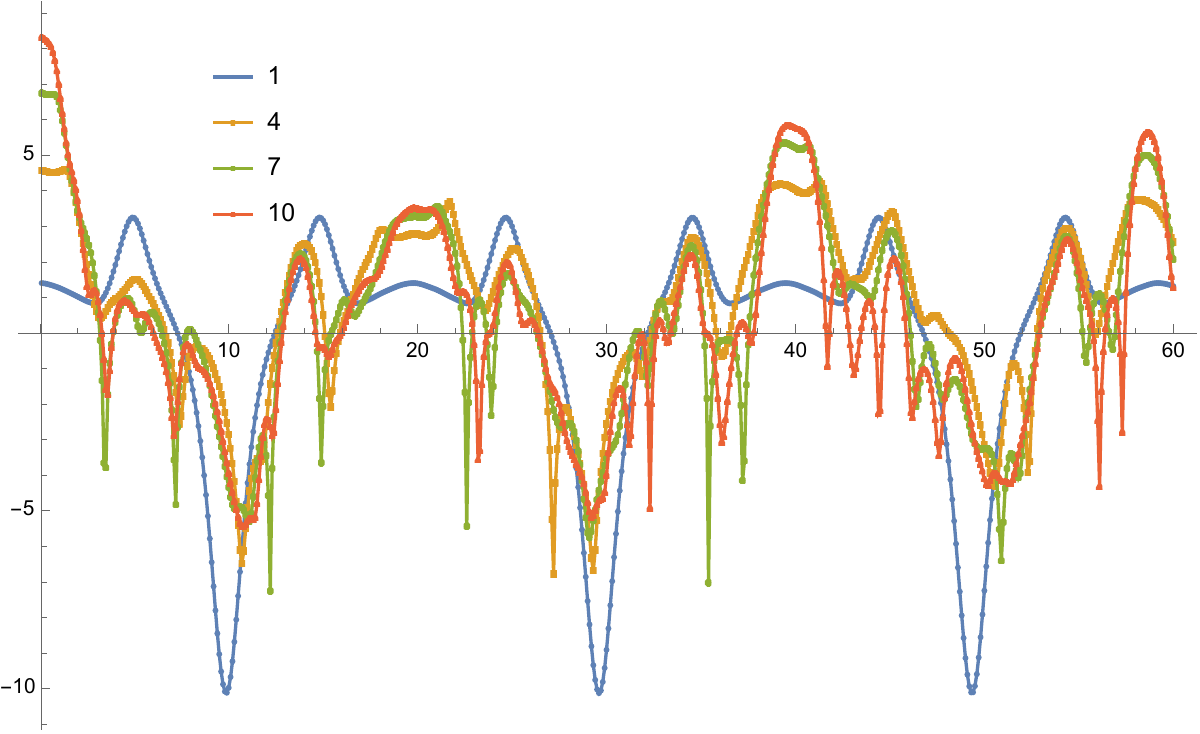}
    \hfill
    \includegraphics[width=5cm]{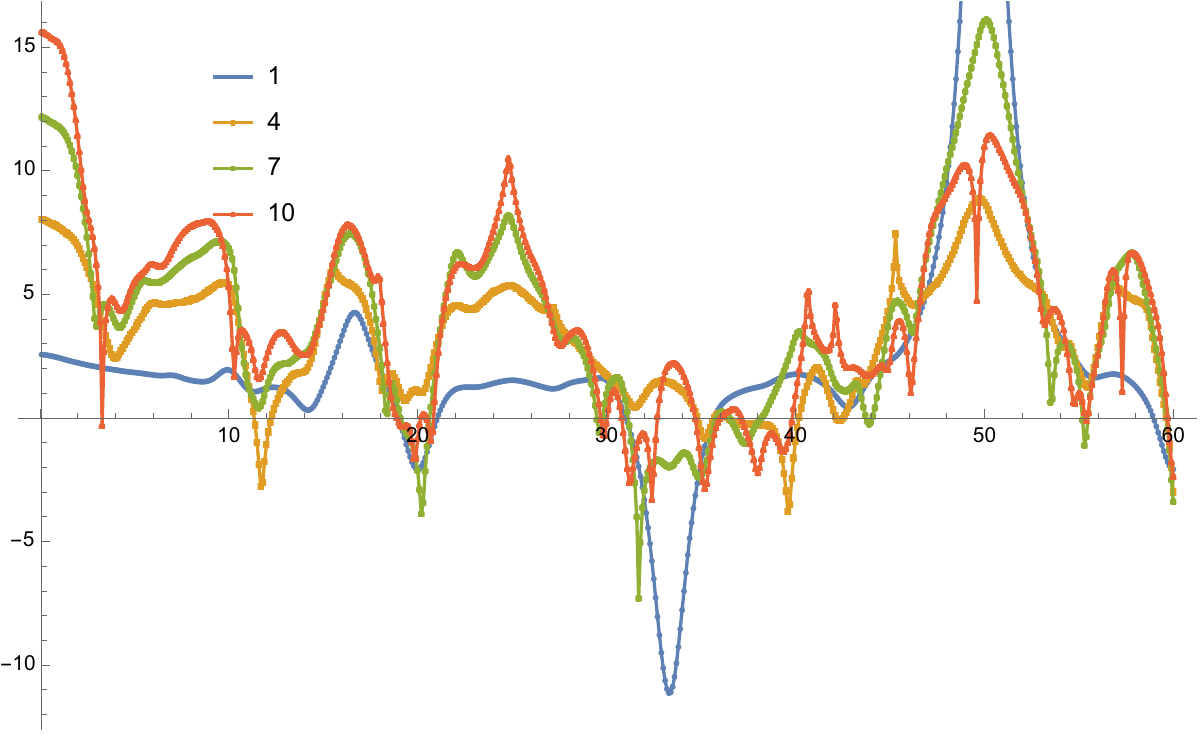}
    \hfill
    \includegraphics[width=5cm]{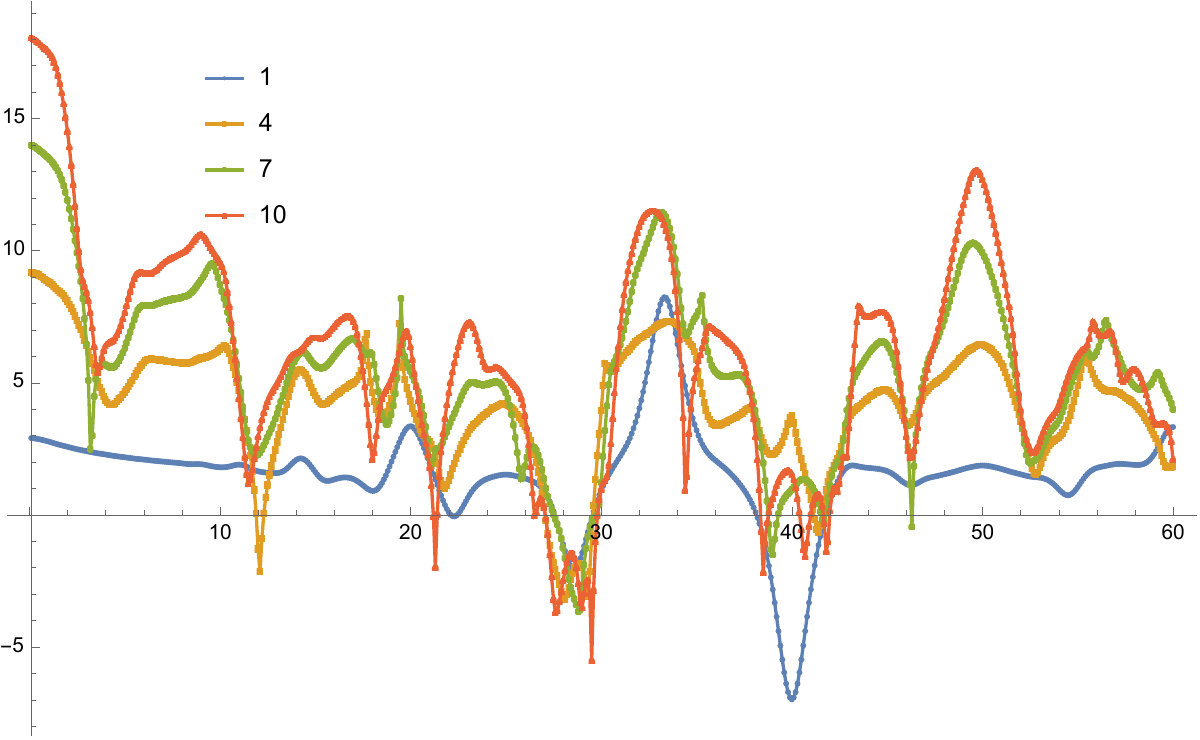}
    \caption{Thermal second R\'enyi pseudo-entropies of compact scalar symmetric orbifolds with increasing $N$, as a function of $\frac{t}{2\pi}$. $\frac{\beta}{2\pi}$ is set to be $1.2$. Left figure is for $R = \pi$, centre for $R = 10$, right for $R = 10\sqrt{2}$.}
    \label{fig:CompScSymRen2_1}
\end{figure}

\begin{figure}[h!]
    \centering
    \includegraphics[width=5cm]{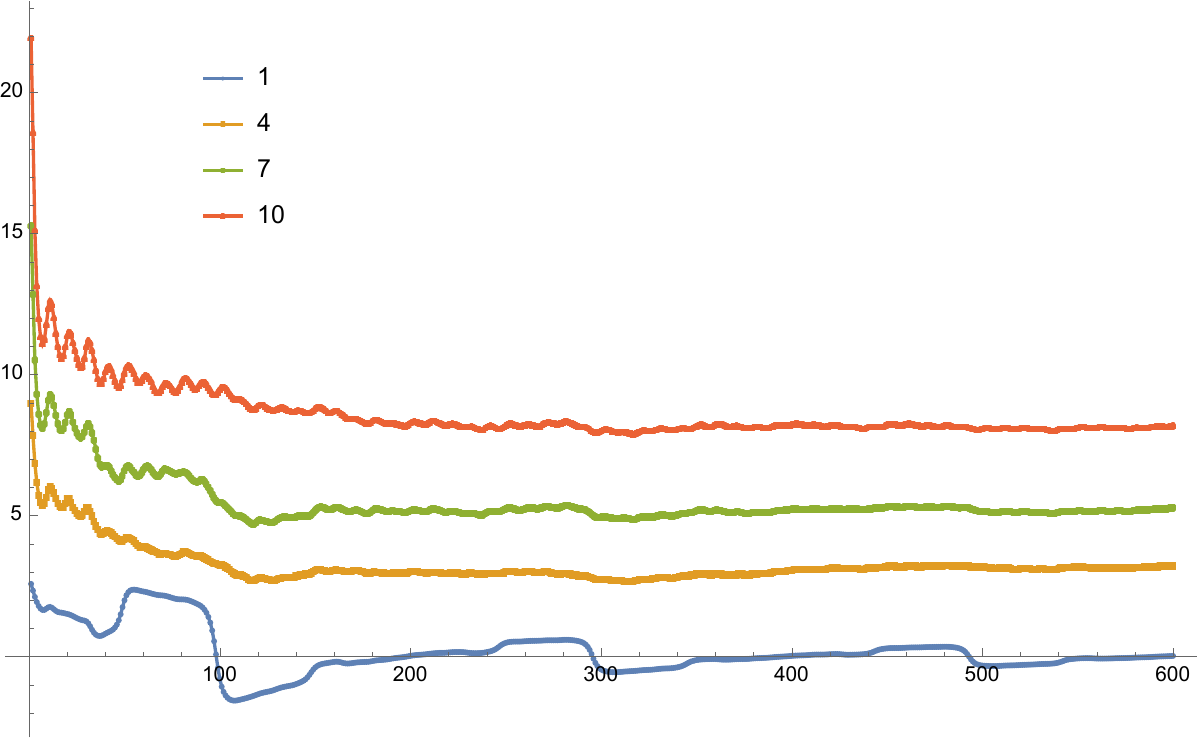}
    \hfill
    \includegraphics[width=5cm]{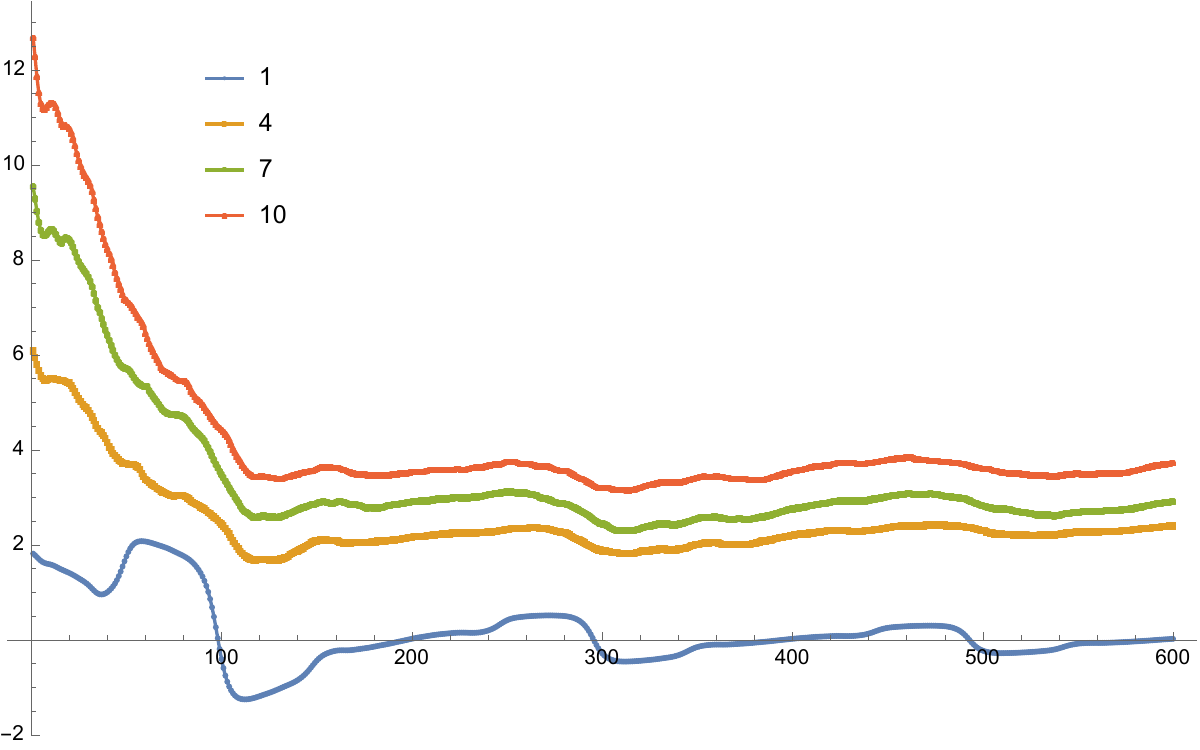}
    \hfill
    \includegraphics[width=5cm]{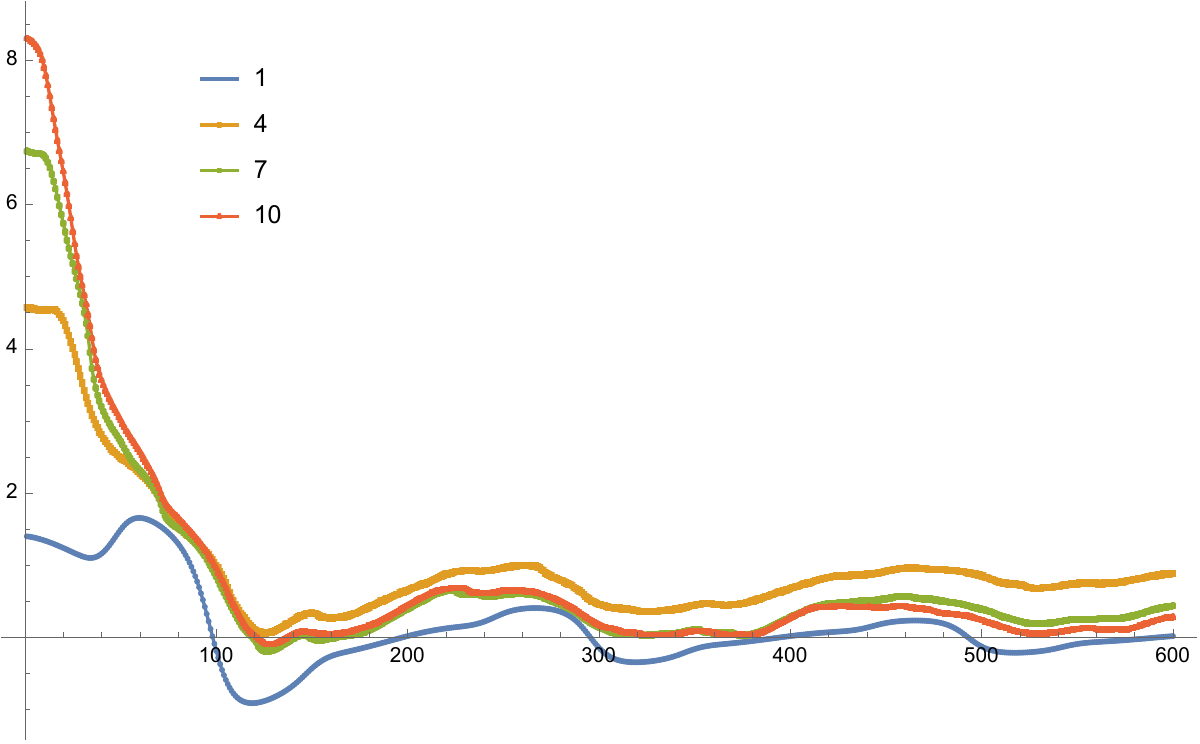}
    \caption{Averaged thermal second R\'enyi pseudo-entropies of compact scalar symmetric orbifolds with increasing $N$, as a function of $\frac{t}{2\pi}$. $R$ is set to be $\pi$. Left figure is for $\frac{\beta}{2\pi} = 0.3$, centre for $0.6$, right for $1.2$. For technical reasons, the time scale is 10 times larger than it should be.}
    \label{fig:CompScSymRen2_2}
\end{figure}

Additional plots of the real part of
the second R\'enyi TPE
are shown in Fig.\,\ref{fig:CompScSymRen2_1}, and plots of the real part of
the second R\'enyi TPE averaged from time 0 to t, given by (\ref{tavepe}), are shown in Fig.\,\ref{fig:CompScSymRen2_2}.
The time $\frac{t}{2\pi}$ is from 0 to 60, and $N$ is set to values of 1, 4, 7, and 10. At the early time region, we can observe the logarithmic behaviour $S(\beta+it)\sim -\frac{N}{2}\log t$. This is also expected from the partition function as it includes a term $Z(\tau)^N$, which is the same as that of $N$ free scalar CFT, as in (\ref{N2SymOr}).

\section{Conclusions and Discussion}\label{sec:Concldis}
In this work we analysed the TPE and its properties in various quantum mechanical and field theoretic setups. This generalisation of the thermal entropy to complex inverse temperatures has several interesting features that we now briefly summarise and point to some future directions. 

Firstly, we saw that the average of the real part of the TPE is directly related to the SFF. This is an interesting avenue that may bridge between quantum chaos (the domain of SFF) and quantum information (where pseudo-entropies originate). Indeed, a lot of work is dedicated to understanding quantum information aspects of pseudo-entropies, such as e.g. its operational meaning or gravity dual. The link we found could potentially shed some of this quantum information light on the SFF and quantum chaos.

On a similar note, we introduced the TPE as a natural step in the pseudo-entropy program \cite{Nakata:2021ubr}, but it also seems very intuitive from the non-equilibrium physics point of view. To our knowledge, the TPE has not appeared in that context before, but partition functions with complex temperatures (as well as general complex couplings) has, and are e.g. important to understand dynamical phases of matter \cite{heyl2013dynamical,wei2014phase,Sarkar:2023pdf}. On the other hand, pseudo-entropies and their excess has already proved useful for diagnosing various phase transitions so it will be interesting to connect these seemingly distant developments in the future. 

One of the main problems on which we made concrete progress with TPE is the relation between its real and imaginary parts provided by the Hilbert transform or the Kramers-Kronig relations. We saw that it holds in all our examples and argued for it based on the decay of generic partition function at large $\beta$ or vanishing of thermal entropy in this limit. It would be very interesting to prove this more rigorously and also check which other, more general classes of, pseudo-entropies \cite{Nakata:2021ubr} can be given by analytic functions with KK relations between their real and imaginary parts. 

The second feature of the TPE that we found in models with continuous spectrum is the $-(1+\gamma)\log(t)$ term. As we saw, it is intimately related to the scaling of the spectral density near the edge. Usually, various observables in this regime of RMT, explore the middle region of the Wigner semi-circle (away from the edges) so it is quite interesting that TPE is sensitive to this information. 

More generally, logarithmic terms, often play a role of indicators of universal properties of the model. Since we found it in RMT and the Schwarzian theory (both paradigmatic examples of the quantum chaotic evolution) it is tempting to relate the $\log(t)$ to chaotic or scrambling properties of the model. However, as we saw in the CFT examples (see summary Table \ref{TableLogt}), this statement should be taken with caution because $\log(t)$ may also arise in the decompactification limit of the integrable compact boson CFT. This suggests that its presence or absence may not be a sufficient condition for quantum chaos or distinguishing scrambling from chaos \cite{Xu:2019lhc,Caputa:2016tgt,Kudler-Flam:2019kxq}. Indeed, we also find that such a logarithmic term is absent in two-dimensional holographic CFTs. Nevertheless, if will be interesting to explore this issue further.\\
Similarly, we observed the $1/2\log(t)$ growth of TPE after the dip in RMT and JT universality class. This is the counterpart of the linear ramp in the SFF and is closely tight with the double cone geometry \cite{Saad:2018bqo,Chen:2023hra}. It will be also important to understand this term better from the holographic perspective and the general proposal for pseudo-entropy \cite{Nakata:2021ubr}.
\begin{table}[h!]
\centering
\begin{tabular}{ | c | c| c | } 
  \hline
  Model: & Coefficient of $-\log(t)$ & $\gamma$ \\ 
  \hline
  Schwarzian & 3/2 & 1/2 \\ 
  \hline
  RMT & 3/2 & 1/2 \\ 
  \hline
  Free Fermion CFT & 0 &  -1 \\ 
  \hline
  Compact scalar CFT, $R<\infty$ & 0 & -1 \\ 
  \hline
  N Compact scalars CFT, $R\to\infty$ & $\frac{N}{2}$ & 
  $\frac{N}{2}-1$ \\ 
  \hline
  Holographic CFT & 0 & -1  \\ 
  \hline
\end{tabular}
  \caption{Summary of the logarithmic term for the models that we considered. The exponent $\gamma$ is defined by (\ref{gammandef}) for a (almost) continuous spectrum. For discrete spectrum, we set $\gamma=-1$, conveniently. Our analysis of  free scalar and fermions orbifolds shows that their logarithmic $t$ terms are the same as for $N$ free scalars and fermions so we omitted them.}\label{TableLogt}
\end{table}

Finally, the connection between averaged TPE and the SFF as well as new examples of SFFs in symmetric orbifold CFTs that we studied naturally align with recent developments on arithmetic chaos, modularity and SFF in rational CFTs \cite{Benjamin:2018kre,Haehl:2023tkr,Haehl:2023xys,Benjamin:2021ygh,Das:2021shw}. Indeed, our numerical calculations of time averaged SFF confirmed the presence of dip and ramp structure for symmetric orbifold CFTs of the free scalar and fermion. The structure of partition functions in orbifold CFTs governed by the elegant mathematics of Hecke operators should not only be amenable to the analysis in these works but could also provide hints on questions about analytical properties of TPE, the KK relations etc. 

\acknowledgments
We would like to thank Diptarka Das, Felix Haehl, Masaki Tezuka and Zixia Wei for useful comments and discussions. We are very grateful to the long-term workshop "Quantum Information, Quantum Matter and Quantum Gravity", YITP-T-23-01, held at YITP, Kyoto U., where this work was initiated. 
P. Caputa and T. Takayanagi also thank the organisers of the program "What is String Theory? Weaving Perspectives Together", held in KITP, UCSB, where a part of this work was developed. This work is supported by MEXT KAKENHI Grant-in-Aid for Transformative Research Areas (A) through the ``Extreme Universe'' collaboration: Grant Number 21H05187. T. Takayanagi is also supported by Inamori Research Institute for Science and by JSPS Grant-in-Aid for Scientific Research (A) No.~21H04469. T. Tsuda is also supported by JSPS KAKENHI Grant Number JP24KJ1374. P. Caputa and B. Chen are supported by NCN Sonata Bis 9 2019/34/E/ST2/00123 grant. B. Chen is supported by the NSF of China through Grant No. 12175238. P. Caputa is supported by the ERC Consolidator grant (number: 101125449/acronym: QComplexity).  Views and opinions expressed are however those of the authors only and do not necessarily reflect those of the European Union or the European Research Council. Neither the European Union nor the granting authority can be held responsible for them.
\appendix
\section{Plots for compact scalar symmetric orbifold CFTs}\label{sec:CompactScalarSymOrbFig}
In this appendix, we present three kinds of plots for compact scalar symmetric orbifold CFTs.
The log-log plots of the SFFs are shown in Fig.\,\ref{fig:CompScSymSFF}, with $t/2\pi \in (0,60)$. We normalise the SFF respectively so that they all start from 1 for $t=0$. $N$ is set to values of 1, 4, 7, 10, 13, and 16.

Plots of the real part of second R\'enyi thermal pseudo-entropy are shown in Fig.\,\ref{fig:CompScSymRen2}, and each of them, averaged from time 0 to t, is shown on Fig.\,\ref{fig:CompScSymRen2Aver}
\begin{align}
\overline{S^{(2)}_{N,S}}\left(\beta+it\right) = \frac{1}{t}\int^t_0 dt' ~ S^{(2)}_{N,S}\left(\beta+it'\right).
\end{align}
The time $t/2\pi$ is from 0 to 60, and $N$ is set to values of 1, 4, 7, and 10.

\begin{landscape}
\begin{figure}
    \centering
    \includegraphics[width=6.5cm]{SymOrbPlots/SFF_RisPi_Bis0.3_Nis1-16_noAver.pdf}
    \hfill
    \includegraphics[width=6.5cm]{SymOrbPlots/SFF_RisPi_Bis0.6_Nis1-16_noAver.pdf}
    \hfill
    \includegraphics[width=6.5cm]{SymOrbPlots/SFF_RisPi_Bis1.2_Nis1-16_noAver.pdf}
\\\vspace{3mm}
    \includegraphics[width=6.5cm]{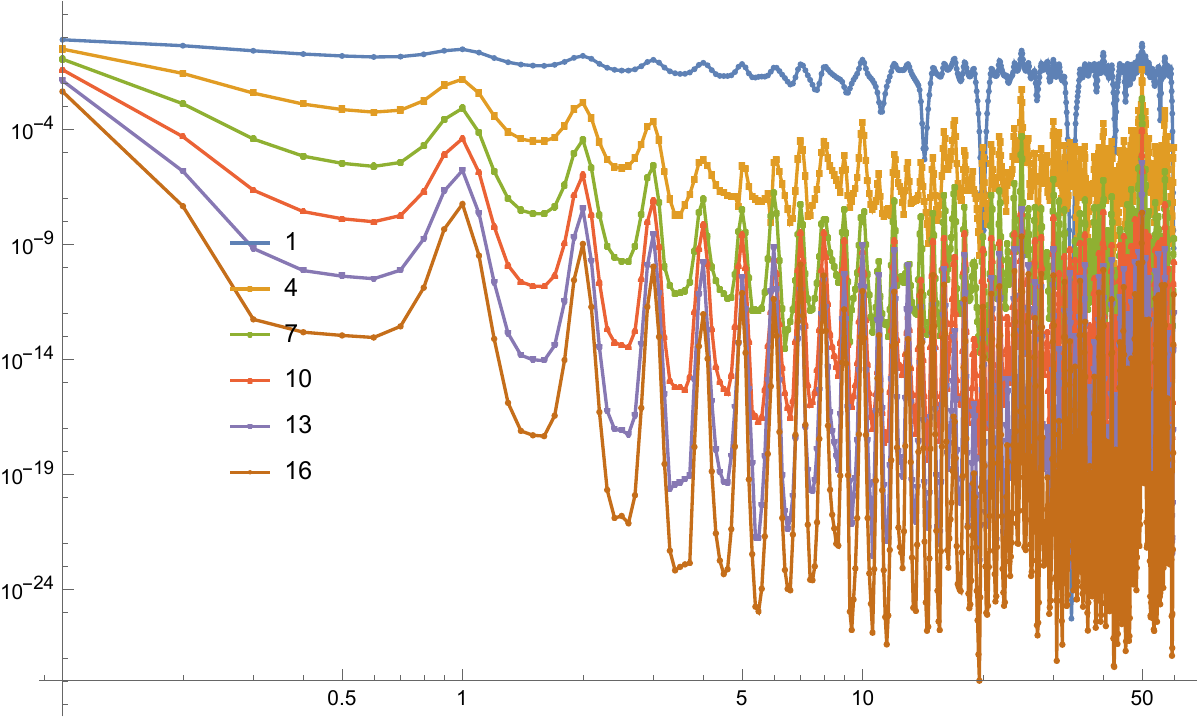}
    \hfill
    \includegraphics[width=6.5cm]{SymOrbPlots/SFF_Ris10_Bis0.6_Nis1-16_noAver.pdf}
    \hfill
    \includegraphics[width=6.5cm]{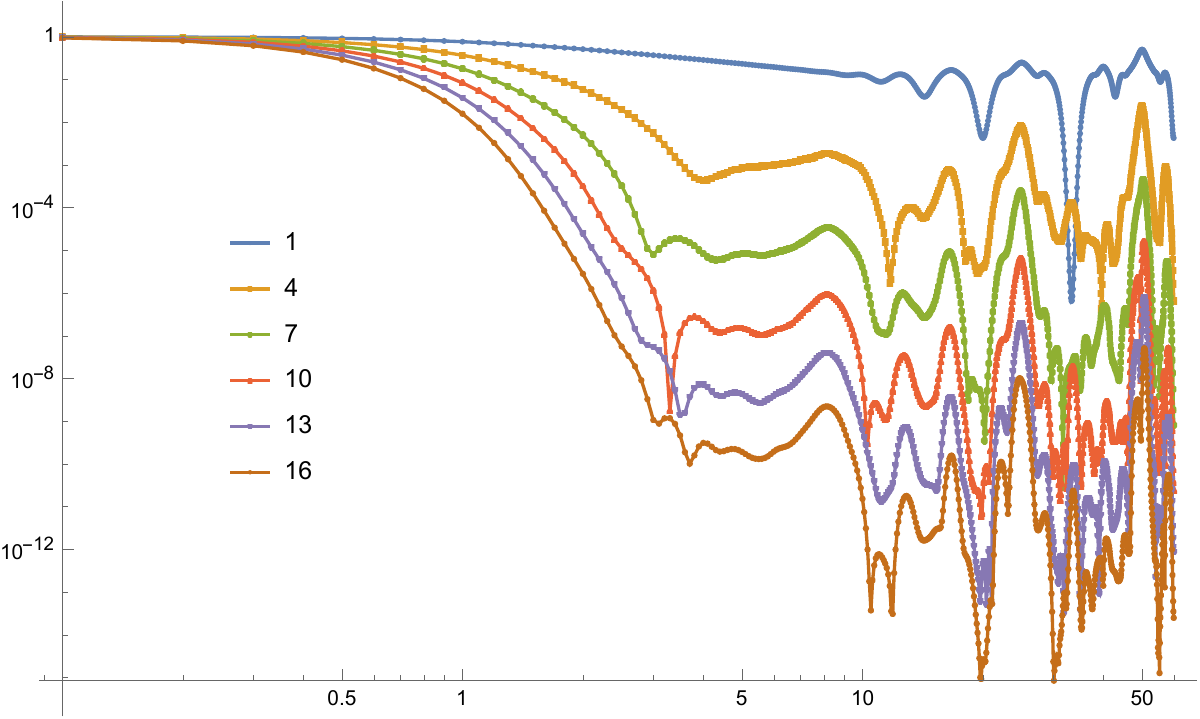}
\\\vspace{3mm}
    \includegraphics[width=6.5cm]{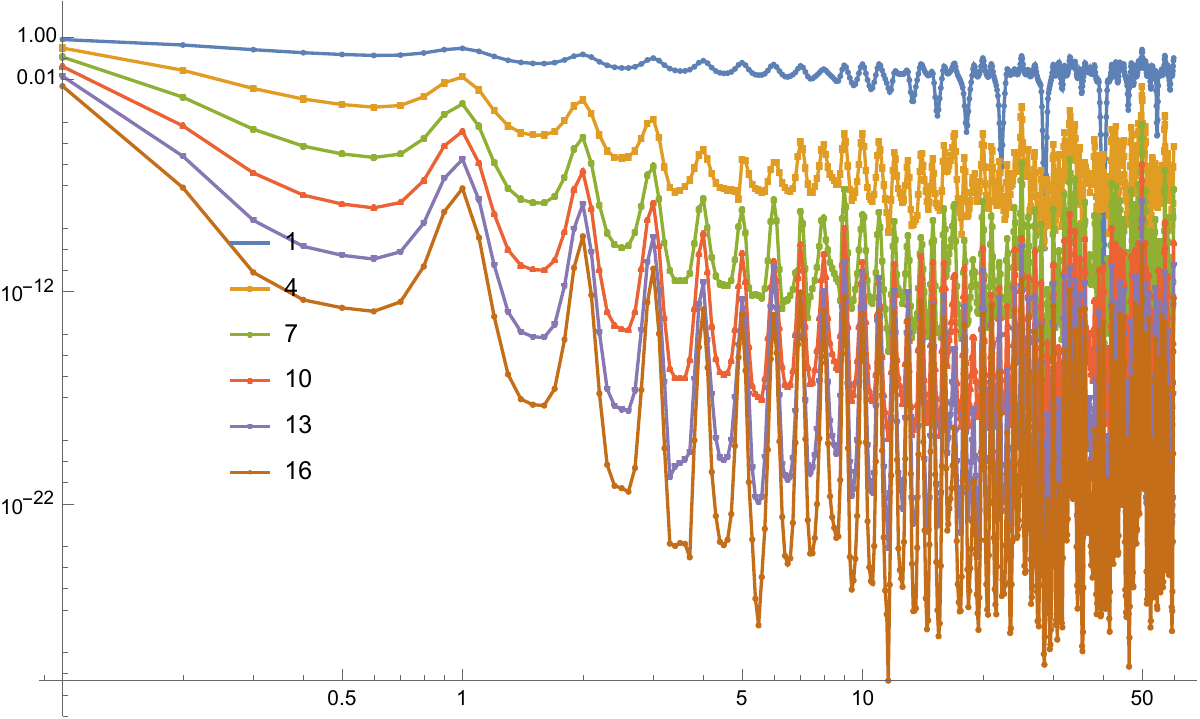}
    \hfill
    \includegraphics[width=6.5cm]{SymOrbPlots/SFF_Ris14_Bis0.6_Nis1-16_noAver.pdf}
    \hfill
    \includegraphics[width=6.5cm]{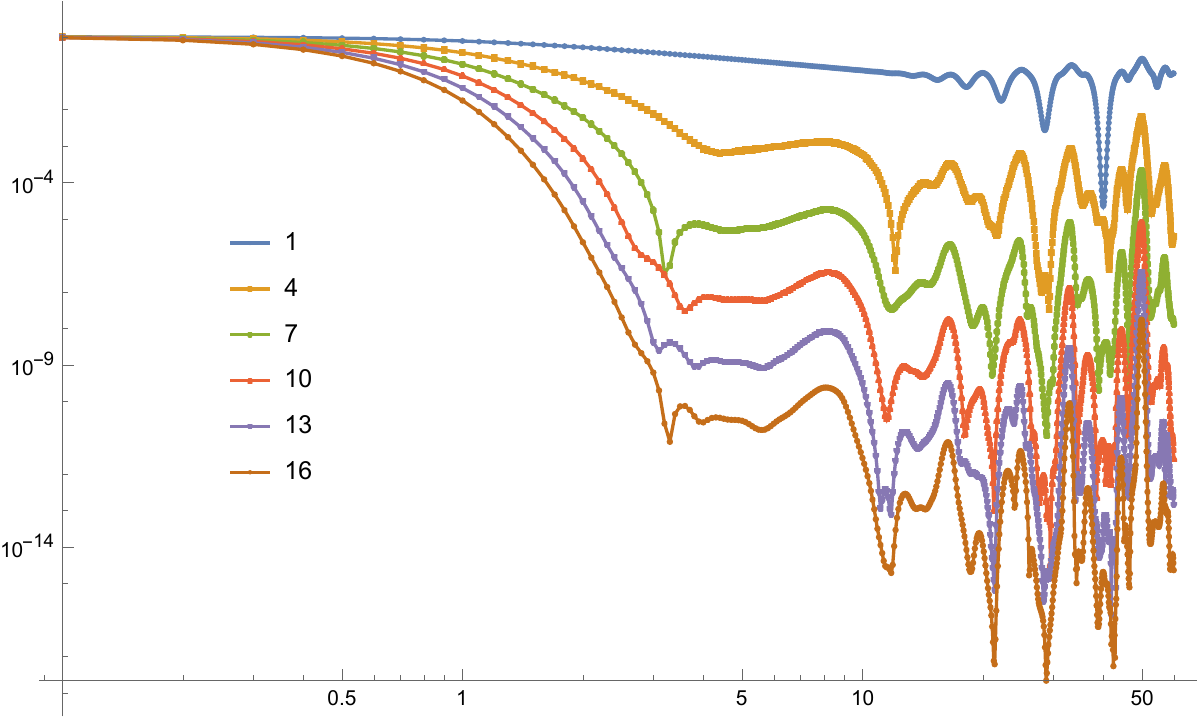}
    \caption{SFFs of compact scalar symmetric orbifolds with increasing $N$. Upper row is for $R = \pi$, middle row is for $R = 10$, lower row is for $R = 10\sqrt{2}$. Left column is $\beta/2\pi = 0.3$, centre column is $\beta/2\pi = 0.6$, right column is $\beta/2\pi = 1.2$.}
    \label{fig:CompScSymSFF}
\end{figure}

\begin{figure}
    \centering
    \includegraphics[width=6.5cm]{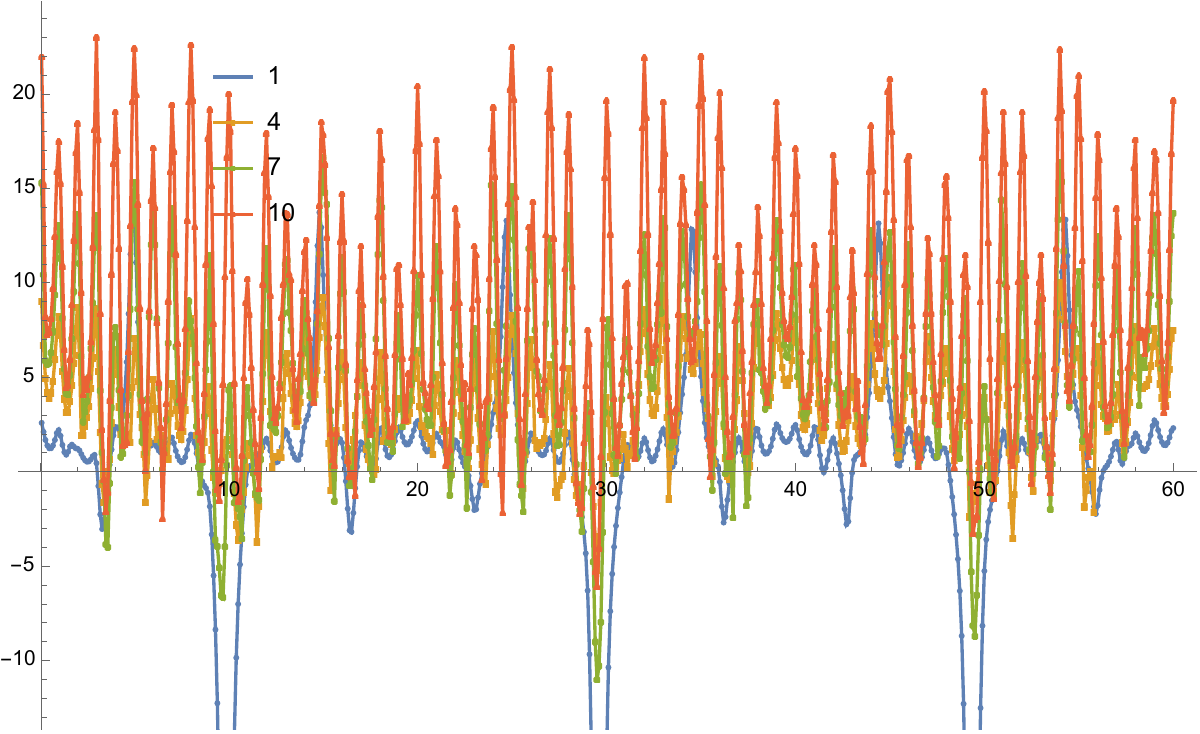}
    \hfill
    \includegraphics[width=6.5cm]{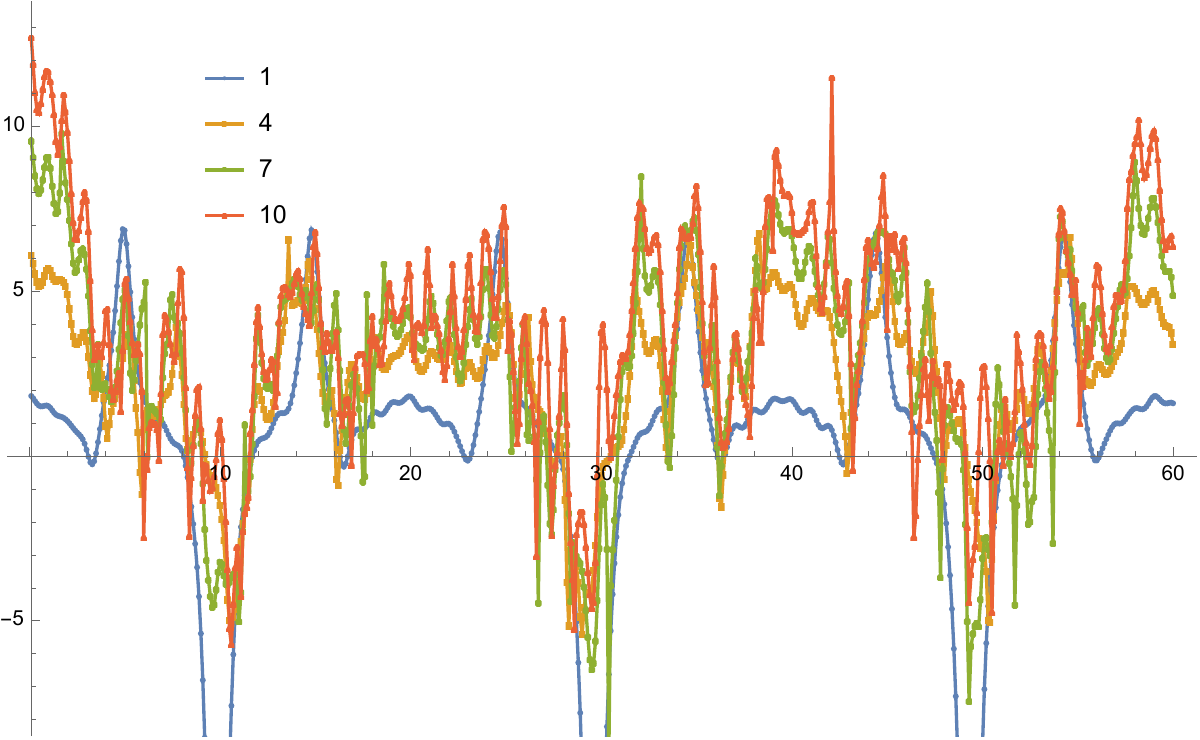}
    \hfill
    \includegraphics[width=6.5cm]{SymOrbPlots/Ren2_RisPi_Bis1.2_Nis1-10_noAver.pdf}
\\\vspace{3mm}
    \includegraphics[width=6.5cm]{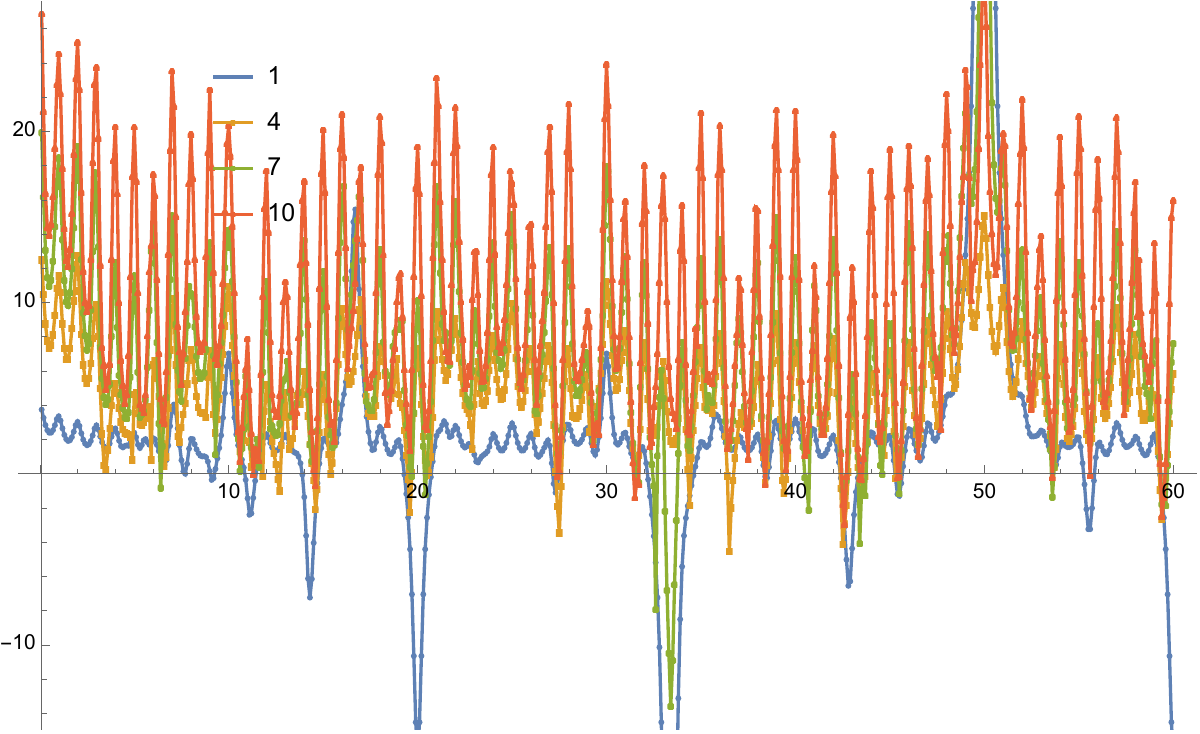}
    \hfill
    \includegraphics[width=6.5cm]{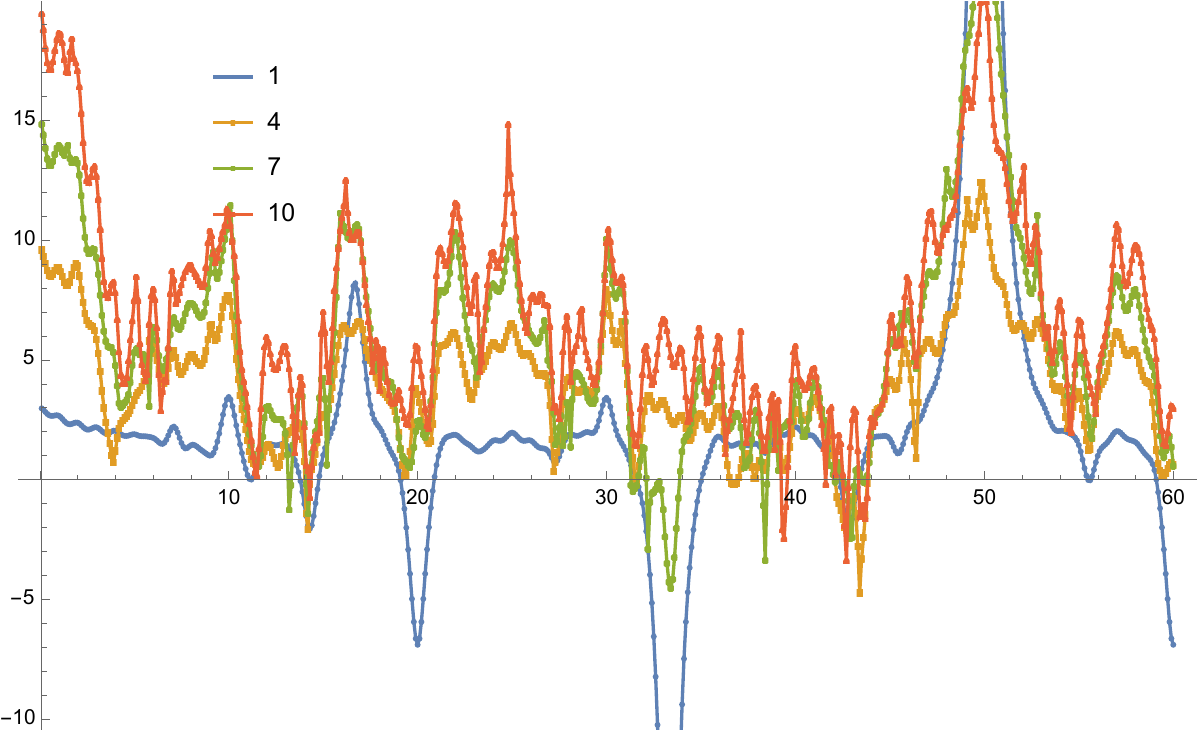}
    \hfill
    \includegraphics[width=6.5cm]{SymOrbPlots/Ren2_Ris10_Bis1.2_Nis1-10_noAver.pdf}
\\\vspace{3mm}
    \includegraphics[width=6.5cm]{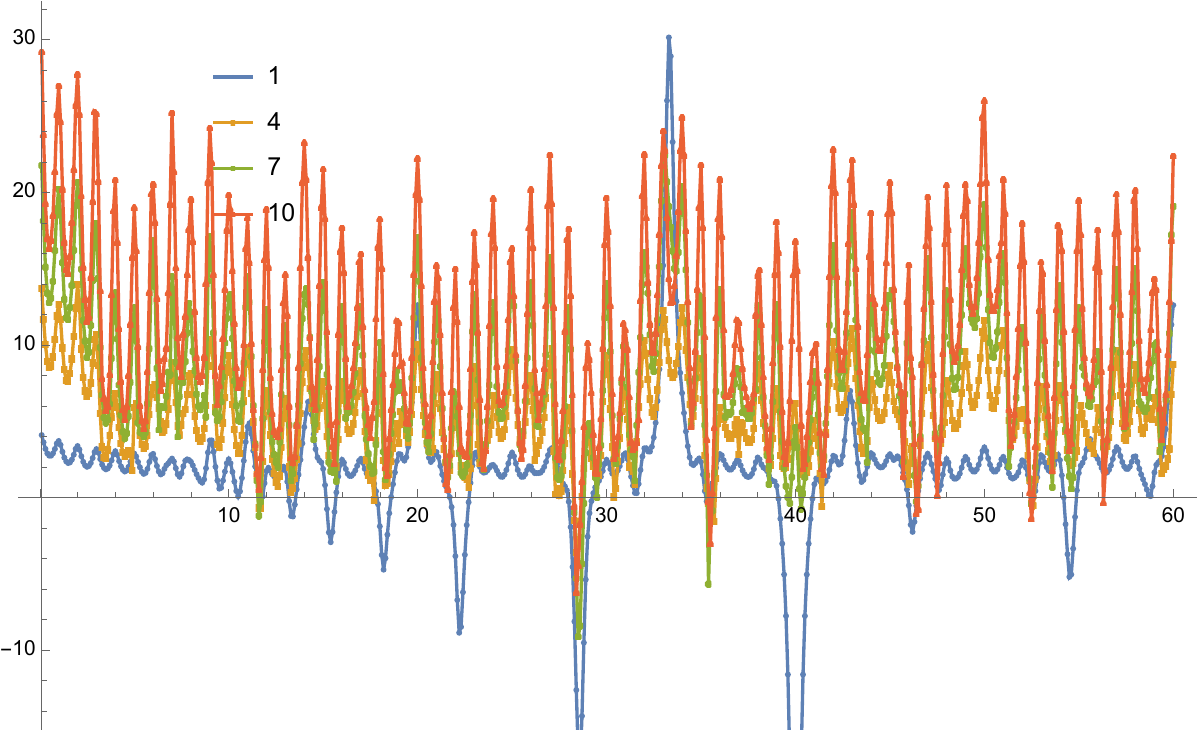}
    \hfill
    \includegraphics[width=6.5cm]{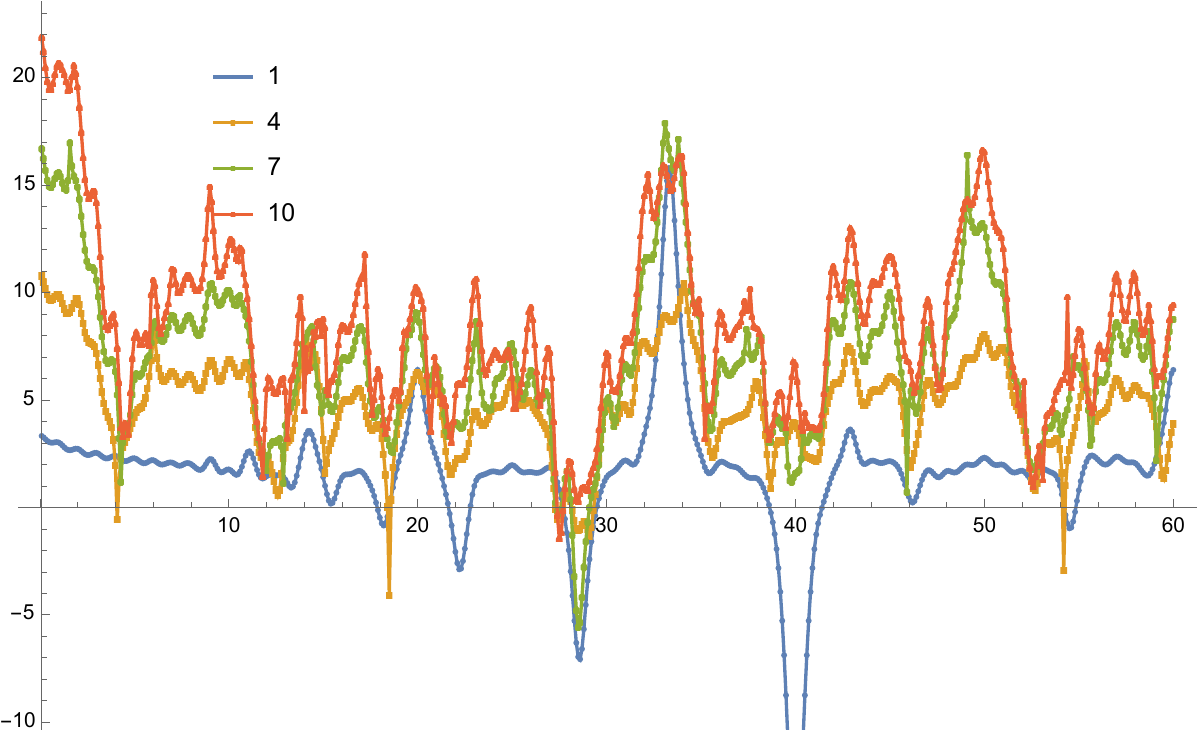}
    \hfill
    \includegraphics[width=6.5cm]{SymOrbPlots/Ren2_Ris14_Bis1.2_Nis1-10_noAver.pdf}
    \caption{Thermal second R\'enyi pseudo-entropies for compact scalar symmetric orbifolds with increasing $N$. Upper row is for $R = \pi$, middle row is for $R = 10$, lower row is for $R = 10\sqrt{2}$. Left column is $\beta/2\pi = 0.3$, centre column is $\beta/2\pi = 0.6$, right column is $\beta/2\pi = 1.2$.}
    \label{fig:CompScSymRen2}
\end{figure}

\begin{figure}
    \centering
    \includegraphics[width=6.5cm]{SymOrbPlots/Ren2_RisPi_Bis0.3_Nis1-10_Aver.pdf}
    \hfill
    \includegraphics[width=6.5cm]{SymOrbPlots/Ren2_RisPi_Bis0.6_Nis1-10_Aver.pdf}
    \hfill
    \includegraphics[width=6.5cm]{SymOrbPlots/Ren2_RisPi_Bis1.2_Nis1-10_Aver.pdf}
\\\vspace{3mm}
    \includegraphics[width=6.5cm]{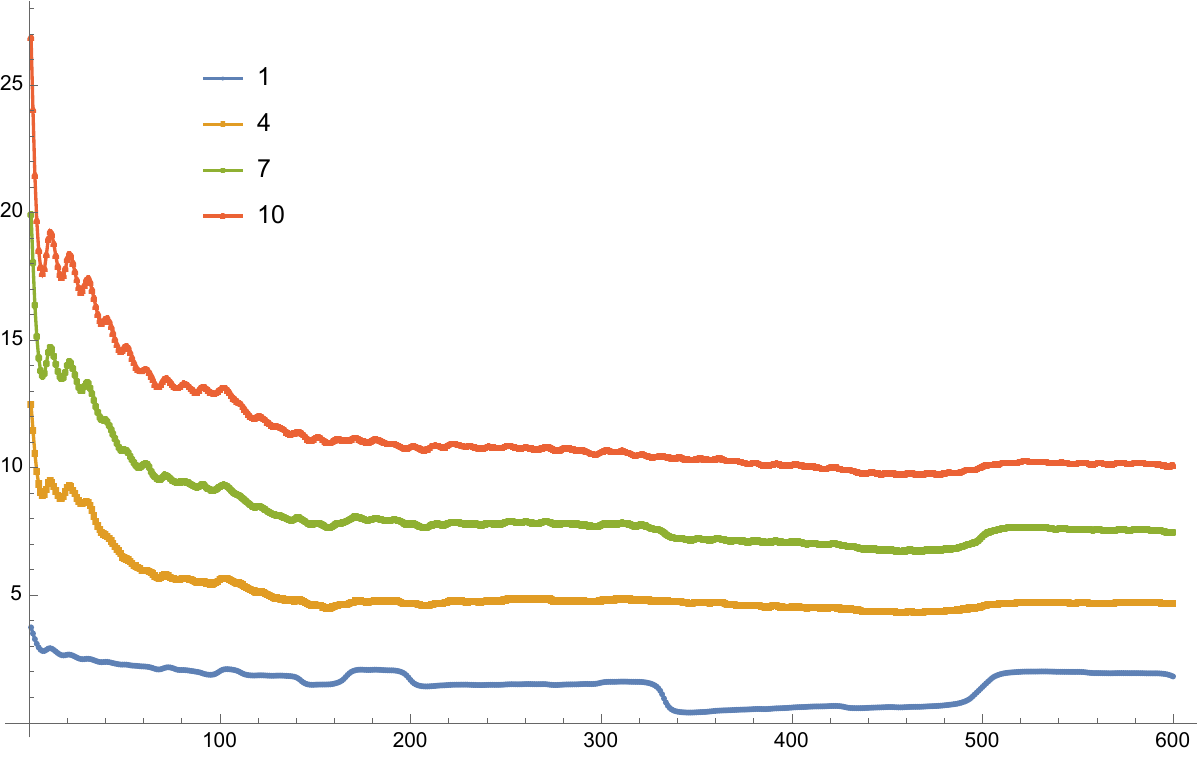}
    \hfill
    \includegraphics[width=6.5cm]{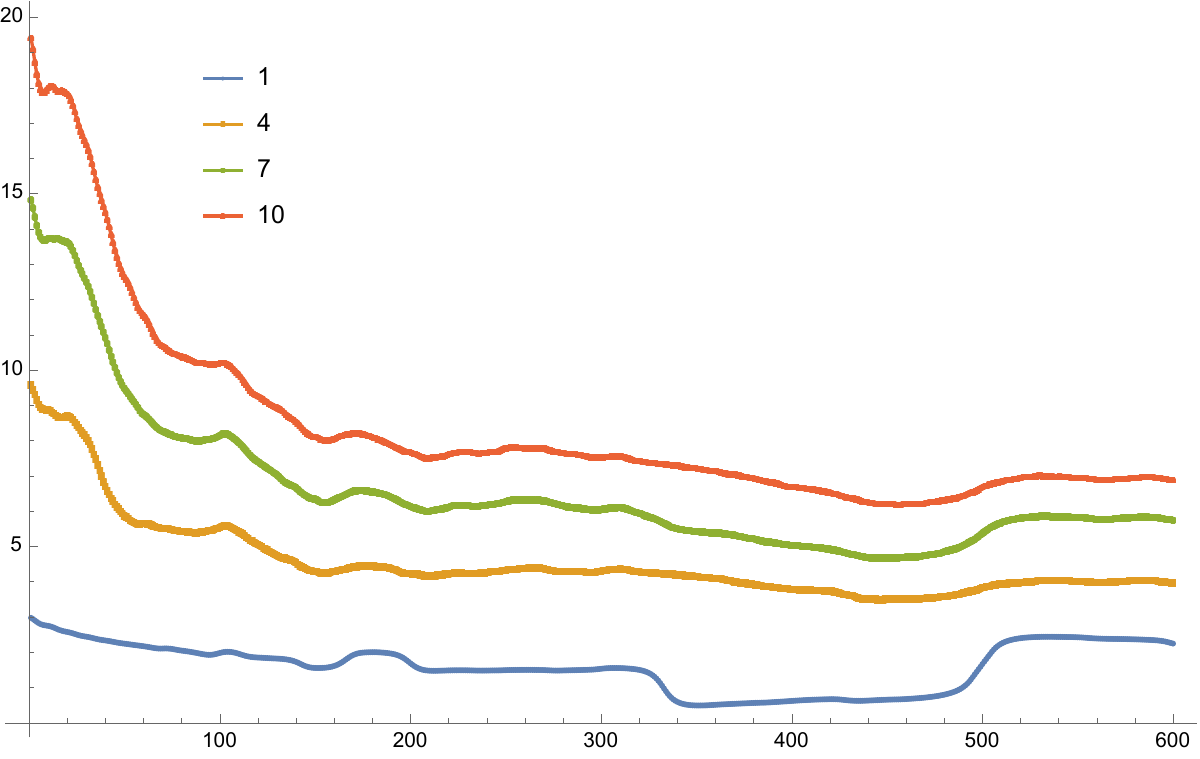}
    \hfill
    \includegraphics[width=6.5cm]{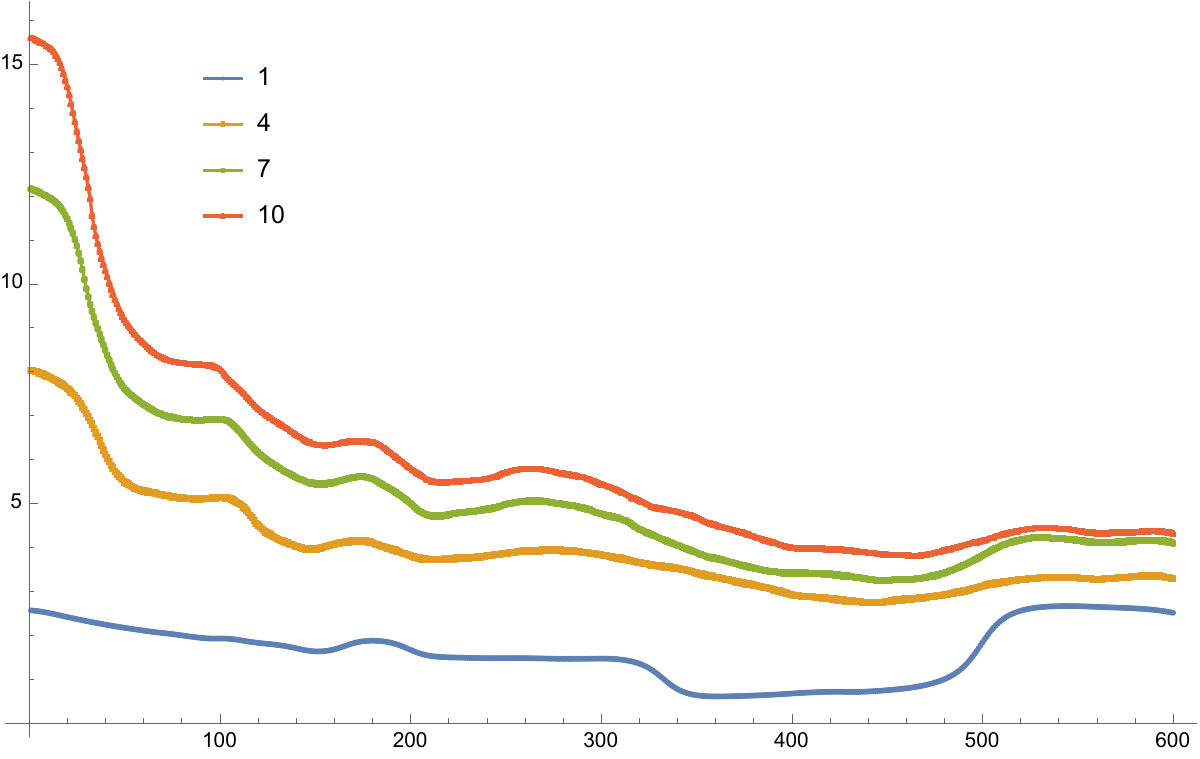}
\\\vspace{3mm}
    \includegraphics[width=6.5cm]{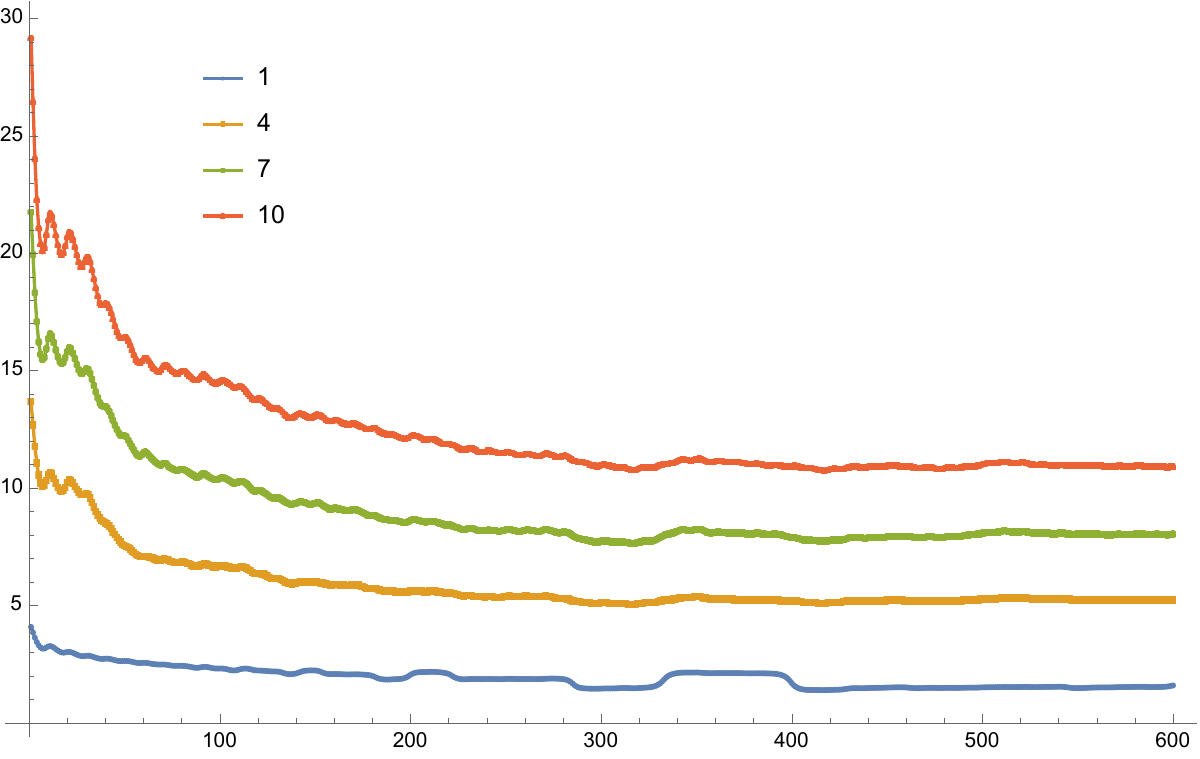}
    \hfill
    \includegraphics[width=6.5cm]{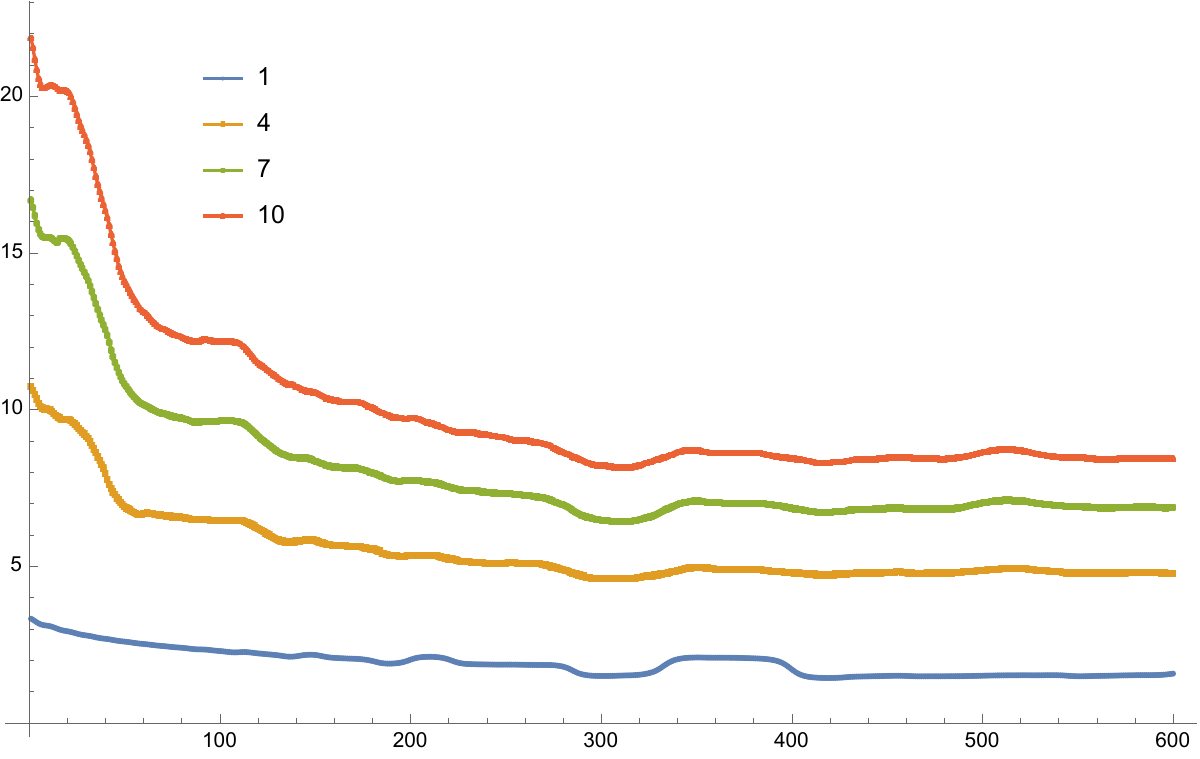}
    \hfill
    \includegraphics[width=6.5cm]{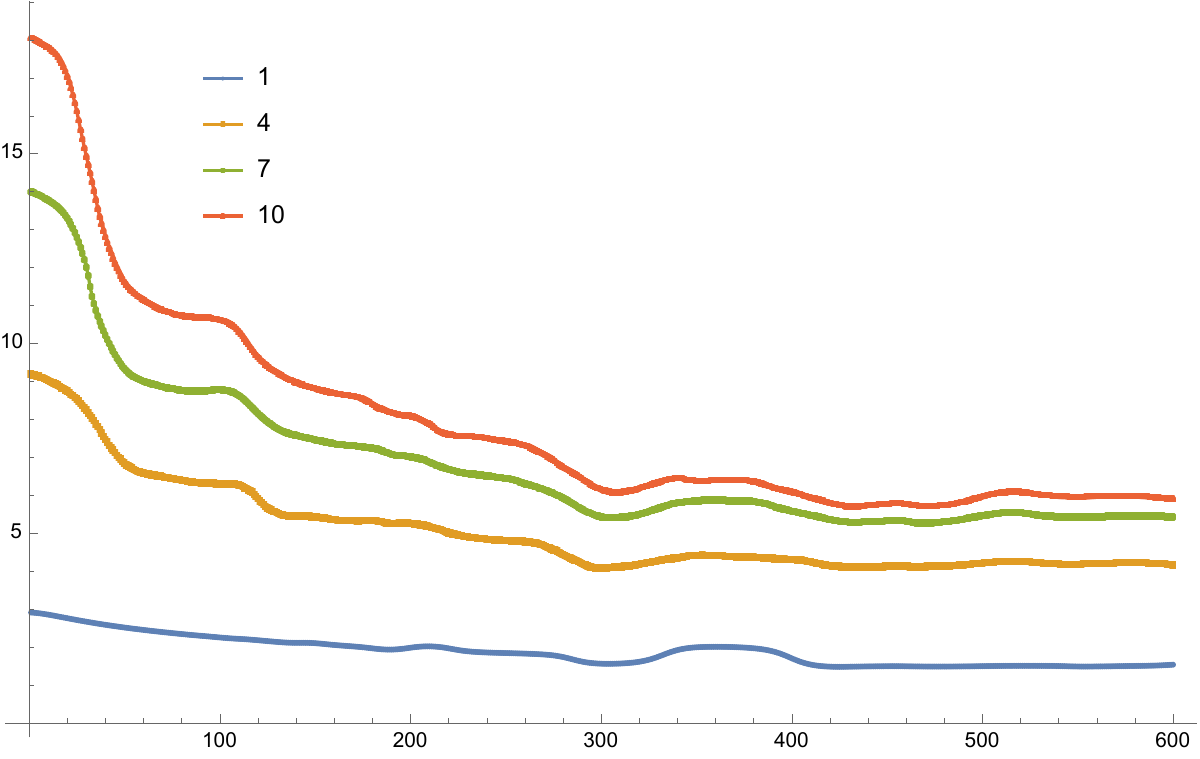}
    \caption{Averaged thermal 2nd R\'enyi pseudo-entropies of compact scalar symmetric orbifolds with increasing $N$. For technical reasons, the time scale is 10 times larger than it should be. Upper row is for $R = \pi$, middle row is for $R = 10$, lower row is for $R = 10\sqrt{2}$. Left column is $\beta/2\pi = 0.3$, centre column is $\beta/2\pi = 0.6$, right column is $\beta/2\pi = 1.2$.}
    \label{fig:CompScSymRen2Aver}
\end{figure}
\end{landscape}

\bibliographystyle{JHEP}
\bibliography{main}

\end{document}